\documentclass[reprint,superscriptaddress,nofootinbib,amsmath,amssymb,aps,onecolumn,notitlepage]{revtex4-1}
\usepackage[a4paper, margin=2.50cm]{geometry}

\usepackage{bbm}
\usepackage{graphicx}
\usepackage{fancyhdr, color}
\usepackage{hyperref}

\newcommand{\nn}{\nonumber\\}

\newcommand{\f}[1]{\mbox{\boldmath$#1$}}

\newcommand{\ord}{{\cal O}}

\newcommand{\ptrace}[2]{{\rm Tr_{#1}}\left\{ #2 \right\}}
\newcommand{\traceB}[1]{{\rm Tr_B}\left\{ #1 \right\}}

\newcommand{\abs}[1]{{\left| #1 \right|}}

\newcommand{\ii}{\mathrm{i}}  

\begin{document}

\fancyhead[C]{\sc \color[rgb]{0.4,0.2,0.9}{Quantum Thermodynamics book}}
\fancyhead[R]{}

\title{The reaction coordinate mapping in quantum thermodynamics}

\author{Ahsan Nazir}
\email{ahsan.nazir@manchester.ac.uk} 
\affiliation{School of Physics and Astronomy, The University of Manchester, Manchester M13 9PL, United Kingdom}

\author{Gernot Schaller}
\email{gernot.schaller@tu-berlin.de} 
\affiliation{Institut f\"ur Theoretische Physik, Technische Universit\"at Berlin, D-10623, Berlin.}

\date{\today}

\begin{abstract}
We present an overview of the reaction coordinate approach to handling strong system-reservoir interactions in quantum thermodynamics. This technique is based on incorporating a collective degree of freedom of the reservoir (the reaction coordinate) into an enlarged system Hamiltonian (the supersystem), which is then treated explicitly. The remaining residual reservoir degrees of freedom are traced out in the usual perturbative manner. The resulting description accurately accounts for strong system-reservoir coupling 
and/or non-Markovian effects over a wide range of parameters, including regimes in which there is a substantial generation of system-reservoir correlations. 
We discuss applications to both discrete stroke and continuously operating heat engines, as well as perspectives for {additional} developments. 
{In particular, we find narrow regimes where strong coupling is not detrimental to the performance of continuously operating heat engines.}
\end{abstract}

\maketitle

\thispagestyle{fancy}

\section{Introduction}


The standard formulation of thermodynamics is based around the assumption of vanishingly weak interactions between the system of interest and any thermal reservoirs to which it is attached. Under these conditions, a system that is put into contact with a single heat reservoir will customarily thermalise to a canonical Gibbs state, with the reservoir itself assumed to remain in thermal equilibrium at a given temperature. A weak coupling approximation is generally very well justified on macroscopic scales, whereby only a small fraction of the system and reservoir constituent degrees of freedom are usually in contact. On nanoscales, however, such a simplification becomes much more questionable, perhaps particularly so when the system and reservoir are quantum mechanical in nature and the generation of non-classical correlations between the two may then play a prominent role.

Several attempts have therefore been made to move beyond weak coupling in this context~\cite{liu2007a,nesi2007b,Horhammer2008,campisi2009a,nicolin2011a,deffner2011a,hausinger2011a,pucci2013a,schaller2013a,ankerhold2014a,iles_smith2014a,gallego2014a,
wang2015a,esposito2015a,esposito2015b,gelbwaser_klimovsky2015a,carrega2015a,strasberg2016a,kac2016a,
cerrillo2016a,seifert2016a,newman2017a,strasberg2017b,miller2017a,mu2017a,jarzynski2017a,freitas2017a,perarnau_llobet2018a}. 
Examples include consideration of the Hamiltonian of mean force, which was reviewed in the previous chapter, 
the hierarchical equations of motion technique to be reviewed in the following chapter, 
and unitary transform methods~\cite{gelbwaser_klimovsky2015a,wang2015a}. Here we consider another {established} but powerful approach~\cite{burkey1984a,garg1985a,martinazzo2011a,woods2014a}
that was recently put forward in a thermodynamic context~\cite{iles_smith2014a,strasberg2016a,newman2017a,schaller2018a,strasberg2018a,restrepo2018a}, namely the reaction coordinate mapping. This technique aims to explicitly account for the most prominent reservoir influences by defining a collective degree of freedom of the reservoir (the reaction coordinate) that is then absorbed into an enlarged supersystem. The remaining reservoir degrees of freedom are then treated in the usual manner as being weakly coupled to the supersystem. As we shall see, one advantage of such a formalism is that it allows us to apply much of the intuition of standard weak-coupling thermodynamics, though now without the restriction to vanishingly weak system-reservoir interactions. Furthermore, dynamical and steady-state benchmarking of the reaction coordinate technique has demonstrated that it is extremely accurate in a number of situations of practical interest, in particular when non-Markovian and strong system-reservoir correlation effects become important~\cite{iles_smith2014a,iles_smith2016a}.

The chapter is organised as follows. In the next section we summarise the main ideas behind the reaction coordinate mapping and give the essential details pertaining to both bosonic and fermionic reservoirs. We then analyse the stationary state of the mapped system, explain how it differs from a Gibbs state of the original unmapped system, and outline its connection to the Hamiltonian of mean force. Subsequently, we discuss applications of the reaction coordinate formalism to discrete stroke heat engines such as the Otto cycle, and then continuously operating thermo-machines through the example of a single-electron transistor. Finally, we present an outlook on potential future applications of the reaction coordinate technique to related problems where accounting for strong system-reservoir coupling is of crucial importance. For completeness, mathematical details 
are presented in appendices.


\section[Reaction Coordinate Mapping]{The reaction coordinate mapping}

{Many typical system-reservoir setups employ simple non-interacting Hamiltonians for the reservoirs and assume that they 
are coupled linearly to the system.
%
This means that the reservoir Hamiltonian is generally quadratic in bosonic or fermionic creation and annihilation operators, 
whereas these operators enter the interaction Hamiltonian only linearly, e.g.~as 
absorption/emission or tunneling terms.
Such systems can be treated with the approach that we shall now discuss.
}

\begin{figure}[ht]
\includegraphics[width=0.45\textwidth,clip=true]{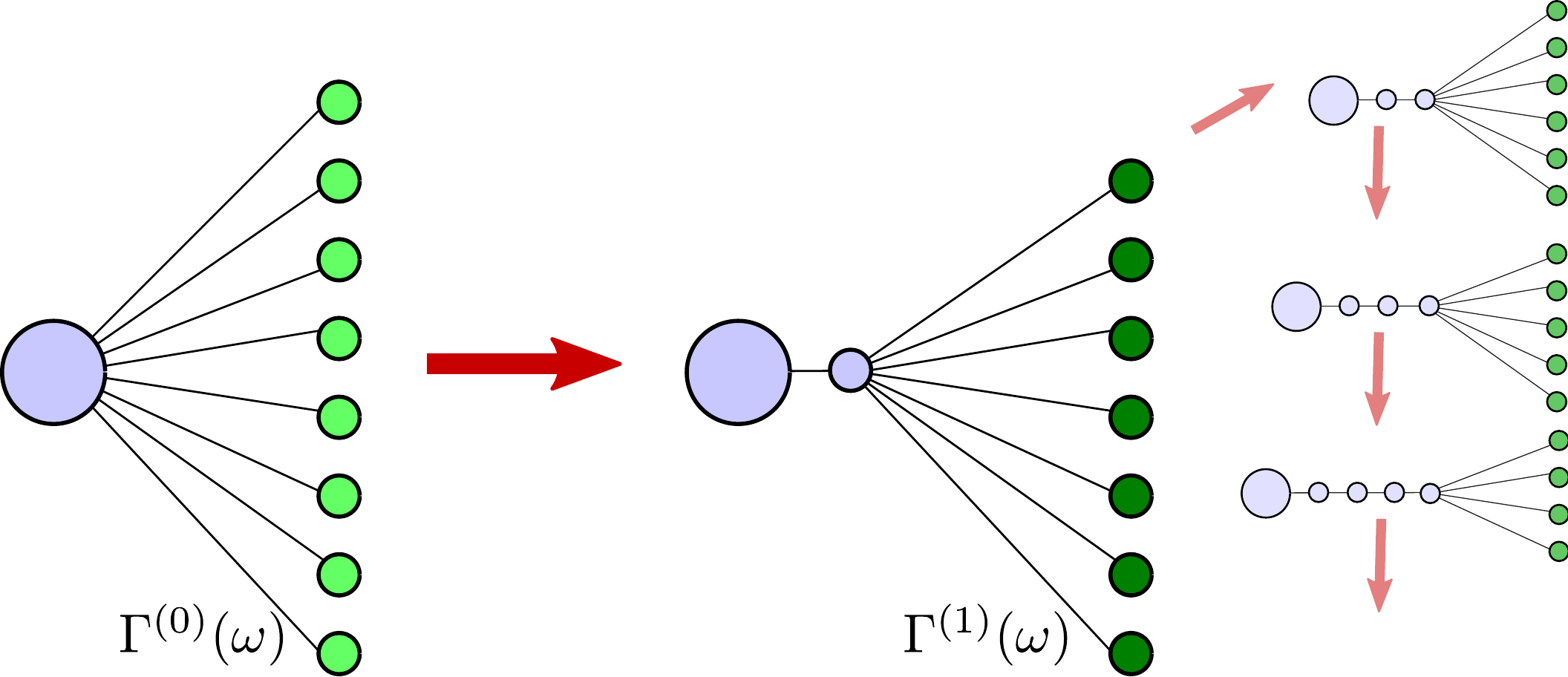}
\caption{\label{FIG:mapping_sketch}
Sketch of the reaction coordinate mapping.
A collective degree of freedom of the reservoir is joined with the system to form the supersystem.
This transforms the spectral density $\Gamma^{(0)}(\omega)$ to $\Gamma^{(1)}(\omega)$.
The mapping can be applied recursively, converting a star configuration into a chain, but also other mappings are conceivable~\cite{huh2014a}.
}
\end{figure}

A schematic of the reaction coordinate mapping is shown in Fig.~\ref{FIG:mapping_sketch}. 
Formally, such mappings can be realized as Bogoliubov transforms, whereby bosonic (or fermionic) annihilation operators $a_k$ are linearly transformed to new 
bosonic (or fermionic) modes $b_q$.
The {\em reaction coordinate} is then selected as one of these new modes $b = b_1$:
\begin{align}\label{EQ:bogoliubov}
a_k = u_{k1} b + \sum_{q>1} u_{kq} b_{q} + v_{k1} b^\dagger + \sum_{q>1} v_{kq} b_q^\dagger\,.
\end{align}
To preserve the bosonic (fermionic anti-) commutation relations, the Bogoliubov transform needs to be symplectic, i.e., 
the matrices formed by the complex-valued coefficients $U=(u_{kq})$ and $V=(v_{kq})$ have to obey the relations
$U U^\dagger \pm V V^\dagger = \f{1}$ and $U V^{\rm T} \pm V U^{\rm T} = \f{0}$, where the $-$ sign accounts for 
bosons and the $+$ sign for fermions, respectively.
For example, when the transform does not mix creation and annihilation operators ($V=\f{0}$), $U$ just needs to be a unitary matrix.
As another example, one can construct a bosonic symplectic transform which mixes between annihilation and creation operators from a given orthogonal transform via 
$u_{kq} = \frac{1}{2} \left(\frac{\bar{a}_k}{\bar{b}_q} + \frac{\bar{b}_q}{\bar{a}_k}\right) \Lambda_{kq}$ and
$v_{kq} = \frac{1}{2} \left(\frac{\bar{a}_k}{\bar{b}_q} - \frac{\bar{b}_q}{\bar{a}_k}\right) \Lambda_{kq}$
where $\bar{a}_k$ and $\bar{b}_q$ are real parameters and the orthogonality relation just imposes the constraint $\sum_q \Lambda_{kq} \Lambda_{k'q} = \delta_{kk'}$.
In general, there is an infinite number of such transformations, and we are interested in the ones giving rise to special mappings
between Hamiltonians as considered below.


\subsection{Spectral bounds}

Many generic bosonic Hamiltonians are specifically designed for the weak-coupling limit.
When these are naively extrapolated towards strong couplings, it may happen that the spectrum of the
complete system is no longer bounded from below, leading to unphysical artifacts.
This problem can be circumvented by writing the initial Hamiltonian as a sum of positive definite terms, which upon expansion 
leads to a renormalized system Hamiltonian.
In our considerations below, we will assume that a decomposition of the total Hamiltonian into completed squares has initially been
performed, such that the spectrum is always bounded from below for any value of the coupling strength.
%


\subsection[Bosonic Reservoirs]{Mappings for continuous bosonic reservoirs}


\subsubsection{Phonon Mapping}

We would like to obtain a map satisfying 
\begin{align}\label{EQ:phononmap}
H &= H_S + S \sum_k \left(h_k a_k + h_k^* a_k^\dagger\right) + \sum_k \omega_k a_k^\dagger a_k\nn
&= H_S + \lambda S (b+b^\dagger) + \Omega b^\dagger b 
+ (b+b^\dagger) \tilde{\sum\limits_k} \left(H_k b_k + H_k^* b_k^\dagger\right)+ \tilde{\sum\limits_k} \Omega_k b_k^\dagger b_k\,,
\end{align}
where $S$ denotes a dimensionless operator acting in the system Hilbert space only, which need not necessarily be bosonic. We term this the phonon mapping due to the position-like couplings both before and after the transformation~\cite{woods2014a}. 
In the second line the symbol $\tilde\sum$ indicates that the mapped reservoir terms contain one less mode (we may not consistently use this
in the following, as the number of modes should become clear from the context).
We further note that we have chosen $\lambda>0$, as a possible phase can be absorbed into the $b$ and $b_k$ operators, eventually only leading to a phase in the $H_k$ coefficients.
With the same argument, we can choose the $h_k$ coefficients to be real-valued from the beginning.
We follow the conventions of defining the corresponding spectral (coupling) densities as
\begin{align}
\Gamma^{(0)}(\omega) = 2\pi \sum_k \abs{h_k}^2 \delta(\omega-\omega_k)\,,\qquad
\Gamma^{(1)}(\omega) = 2\pi \tilde{\sum\limits_k} \abs{H_k}^2 \delta(\omega-\Omega_k)\,.
\end{align}
In our units, these have dimensions of energy.

In terms of Bogoliubov transforms, the mapping~(\ref{EQ:phononmap}) can be realized with a normal-mode transformation, where
\begin{align}
u_{kq} &= \frac{1}{2} \left(\sqrt{\frac{\omega_k}{\Omega_q}} + \sqrt{\frac{\Omega_q}{\omega_k}}\right) \Lambda_{kq}\,,\qquad
v_{kq} = \frac{1}{2} \left(\sqrt{\frac{\omega_k}{\Omega_q}} - \sqrt{\frac{\Omega_q}{\omega_k}}\right) \Lambda_{kq}\,,
\end{align}
with real-valued orthogonal matrix $\Lambda_{kq}$, and where the $\omega_k$ are the natural frequencies of the original modes $a_k$ and 
the $\Omega_q$ the energies of the transformed modes; specifically $\Omega_1=\Omega$ is the reaction coordinate energy.
Some algebra shows that the first column of the orthogonal matrix is fixed by the original system-reservoir coupling
$\Lambda_{k1} = \frac{h_k}{\lambda} \sqrt{\frac{\omega_k}{\Omega}}$.
With this, one obtains the new coupling strength from the old spectral density via
\begin{align}\label{EQ:newphonon1}
\lambda^2 &= \frac{1}{2\pi \Omega} \int_0^\infty \omega \Gamma^{(0)}(\omega) d\omega\,,
\end{align}
where the energy of the reaction coordinate is determined by
\begin{align}\label{EQ:newphonon2}
\Omega^2 = \frac{\int_0^\infty \omega^3 \Gamma^{(0)}(\omega) d\omega}{\int_0^\infty \omega \Gamma^{(0)}(\omega) d\omega}\,,
\end{align}
such that both $\lambda$ and $\Omega$ have dimensions of energy, see Refs.~\cite{martinazzo2011a,strasberg2016a} 
(note the different factor of two in the definition of the spectral densities).
%

With the mapping~(\ref{EQ:phononmap}), manipulations on the Heisenberg equations of motion (see Appendix~\ref{APP:phonon_phonon}) tell us that under 
appropriate conditions the transformed spectral density can be obtained from the old one by the following transformation:
\begin{align}\label{EQ:phonon3}
\Gamma^{(1)}(\omega) &= \frac{4\lambda^2 \Gamma^{(0)}(\omega)}{\left[\frac{1}{\pi}{\cal P} \int \frac{\Gamma^{(0)}(\omega')}{\omega'-\omega} d\omega'\right]^2 + \left[\Gamma^{(0)}(\omega)\right]^2}\,,
\end{align}
where ${\cal P}$ denotes the Cauchy principal value, and we see that $\Gamma^{(1)}(\omega)$ also has units of energy.
In this relation, the analytic continuation of the spectral density to the complete real axis $\Gamma(-\omega) = -\Gamma(+\omega)$ is understood.
This is again compatible with previous discussions~\cite{martinazzo2011a} when the factor of two in the definition of the spectral densities
and the correct dimensionality of coupling coefficients in each representation are taken into account.
Further, the mapping can be applied recursively, e.g. in the next step we have $S=b+b^\dagger$, 
and its convergence properties have been thoroughly investigated~\cite{martinazzo2011a,woods2014a}.
It is straightforward to see that when rescaling $\Gamma^{(0)}(\omega) \to \alpha \Gamma^{(0)}(\omega)$ with some dimensionless constant $\alpha>0$, 
we will just modify the transformed system-reaction coordinate coupling $\lambda \to \sqrt{\alpha} \lambda$, but the residual spectral density $\Gamma^{(1)}(\omega)$ will remain unaffected.
This already tells us that the method can be used to explore the strong-coupling limit of extremely large $\alpha$ whenever the residual coupling is small.
We furthermore see that the spectrum of the supersystem Hamiltonian is bounded from below for any value of $\lambda$ when one
can decompose $H_S = H_S^0 + \frac{\lambda^2}{\Omega} S^2$, where 
$H_S^0$ is bounded from below.

\begin{figure}[t]
\begin{tabular}{cc}
\includegraphics[width=0.45\textwidth,clip=true]{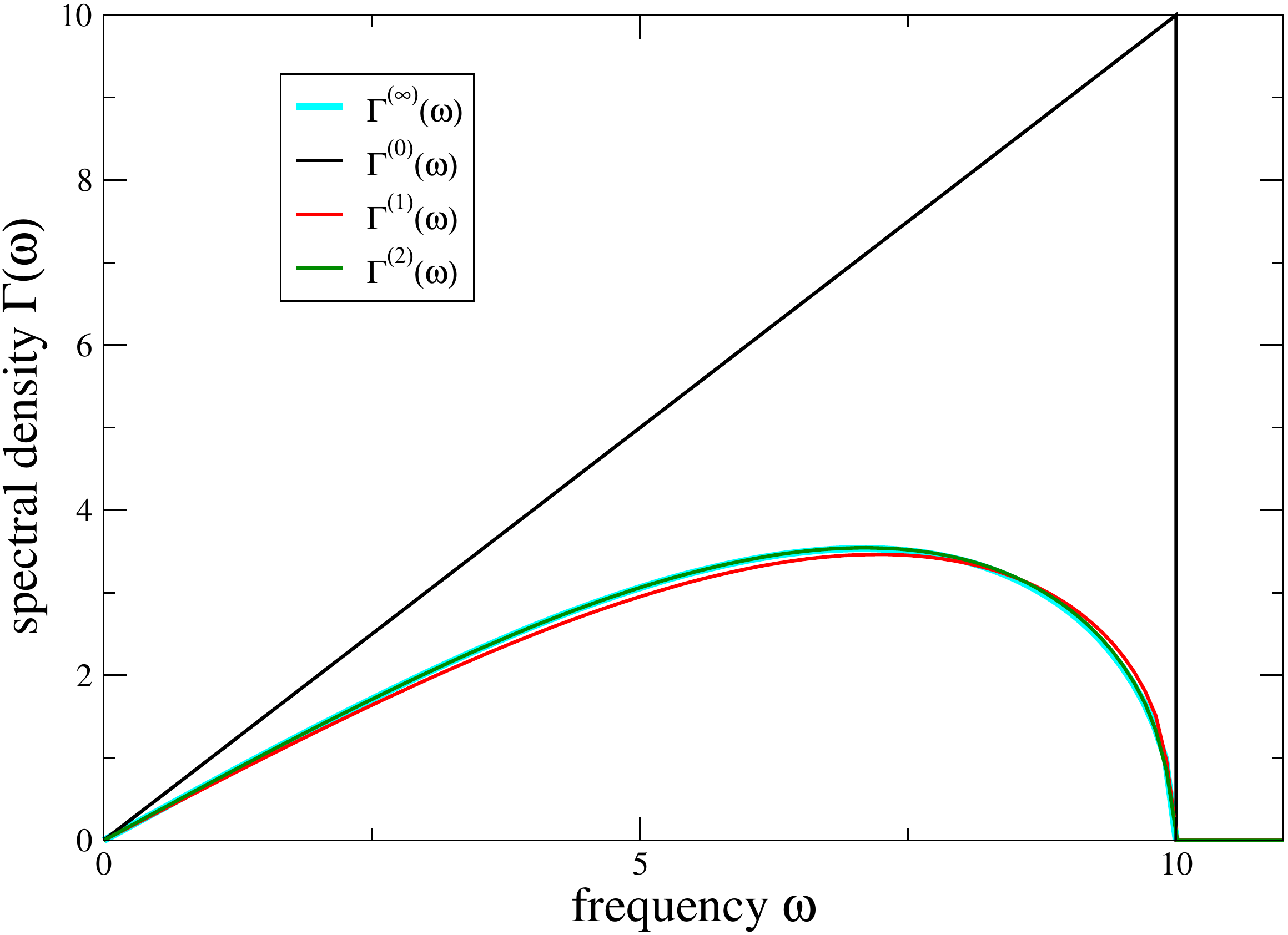} &
\includegraphics[width=0.45\textwidth,clip=true]{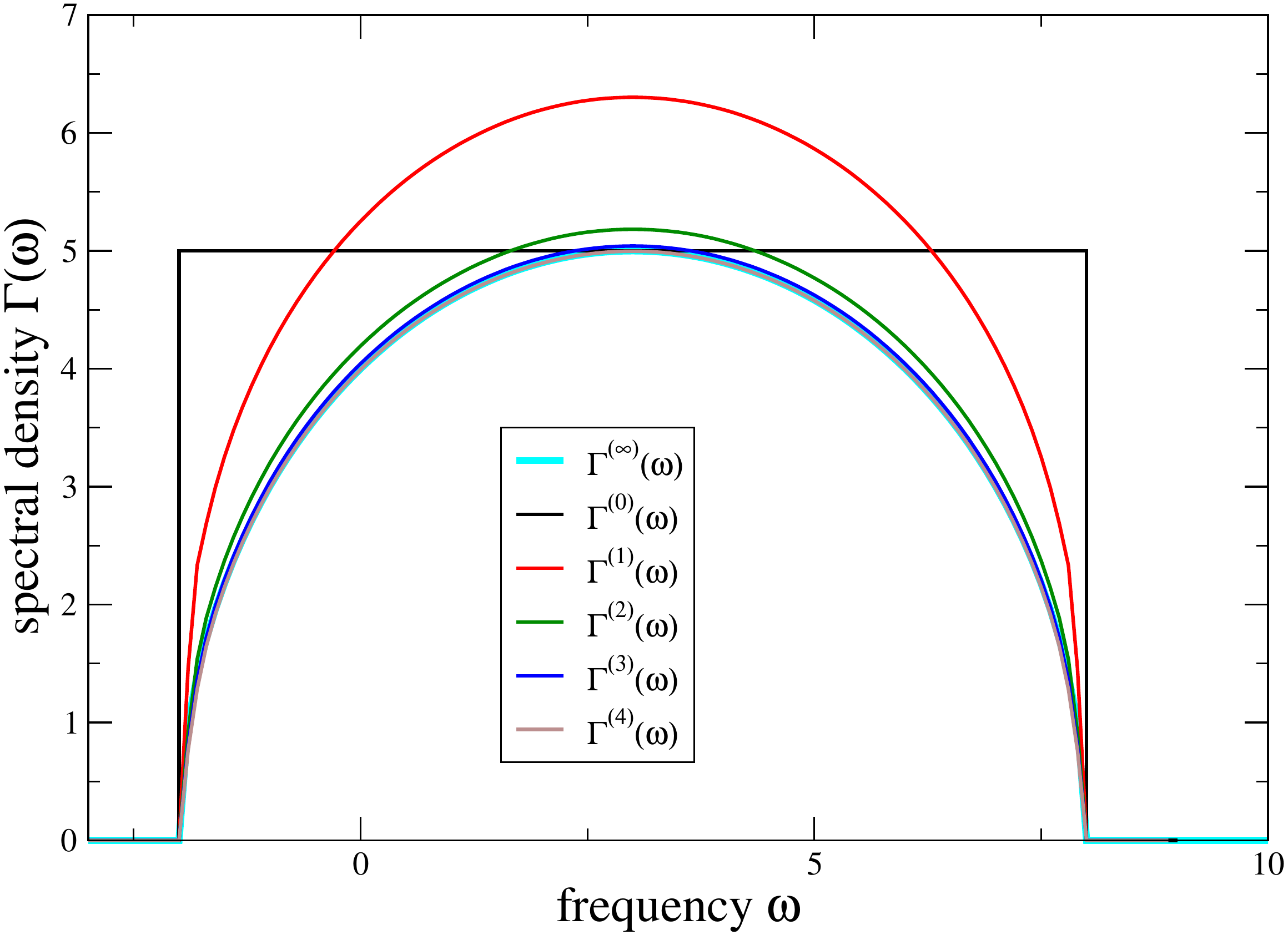}
\end{tabular}
\caption{\label{FIG:convergence}
{\bf Left:} Numerical bosonic phonon mapping of an initially linear spectral density with a rigid cutoff at $\omega_{\rm ct}=10$.
The light blue curve in the background corresponds to the analytic Rubin spectral density in the bottom right of Table~\ref{TAB:spectral_densities_phonon}.
{\bf Right:}
Numerical fermionic particle mapping of an initially box-shaped spectral density with rigid cutoffs at $\omega_{\rm ct,1}=-2$ and 
$\omega_{\rm ct,2}=8$.
The light blue curve in the background corresponds to the semicircle spectral density in the bottom right of Table~\ref{TAB:spectral_densities_fermion} ($\delta=5$ and $\epsilon=3$).
}
\end{figure}

In Table~\ref{TAB:spectral_densities_phonon} we summarize a few spectral densities for which an analytic computation of the mapping relations 
according to Eqs.~(\ref{EQ:newphonon1}),~(\ref{EQ:newphonon2}), and~(\ref{EQ:phonon3}) is possible. One can see that without a rigid cutoff, one may soon obtain spectral densities that do not decay fast enough to allow for another recursive mapping.
With a rigid cutoff, we can however observe convergence towards a stationary Rubin-type spectral density
as in the bottom right of Table~\ref{TAB:spectral_densities_phonon}, see also Fig.~\ref{FIG:convergence} (left panel).

\begin{table}[t]
\begin{tabular}{|c|c|c|c|}
\hline
$\Gamma^{(0)}(\omega)$ & $\lambda^2$ & $\Omega$ & $\Gamma^{(1)}(\omega)$\\
\hline
\hline
$\frac{8 \Gamma \delta^4 \epsilon \omega\left(\omega^2+\delta^2+\epsilon^2\right)}{\left[\delta^2+(\omega-\epsilon)^2\right]^2
\left[\delta^2+(\omega+\epsilon)^2\right]^2}$ & 
$\frac{\Gamma  \delta \epsilon }{4 \sqrt{3 \delta^2+\epsilon^2}}$ & 
$\sqrt{3 \delta^2+\epsilon^2}$ &
$\frac{8 \delta^3 \omega  \left(\delta^2+\omega^2+\epsilon^2\right)}{\sqrt{3 \delta^2+\epsilon^2} \left[2 \omega^2 \left(5 \delta^2-\epsilon^2\right)+\left(3 \delta^2+\epsilon^2\right)^2+\omega^4\right]}$\\
\hline
$\frac{4 \Gamma \delta^5 \omega^3}{\left[\delta^2+(\omega-\epsilon)^2\right]^2\left[\delta^2+(\omega+\epsilon)^2\right]^2}$ &
$\frac{\Gamma \delta^2}{16 \sqrt{5 \delta^2+\epsilon^2}}$ & 
$\sqrt{5\delta^2+\epsilon^2}$ &
$\frac{16 \delta^3 \omega^3}{\sqrt{5 \delta^2+\epsilon^2} \left[\delta^4+2 \delta^2 \left(7 \omega^2+\epsilon^2\right)+\left(\omega^2-\epsilon^2\right)^2\right]}$\\
\hline
$\Gamma \frac{\omega}{\omega_{\rm m}} \Theta(\omega_{\rm m}-\omega)$ &
$\frac{1}{6\pi} \sqrt{\frac{5}{3}} \Gamma \omega_{\rm m}$ &
$\sqrt{\frac{3}{5}} \omega_{\rm m}$ &
$\frac{2 \sqrt{\frac{5}{3}} \pi  \omega  \omega_{\rm m}^2}{3 \left[\pi^2 \omega^2+4 \omega {\rm arctanh}\left(\frac{\omega}{\omega_{\rm m}}\right) 
\left(\omega {\rm arctanh}\left(\frac{\omega}{\omega_{\rm m}}\right)-2 \omega_{\rm m}\right)+4 \omega_{\rm m}^2\right]}$\\
\hline
$\Gamma \frac{\omega}{\omega_{\rm m}} \sqrt{1-\frac{\omega^2}{\omega_{\rm m}^2}} \Theta(\omega_{\rm m}-\omega)$ & 
$\frac{\Gamma  \omega_{\rm m}}{16 \sqrt{2}}$ & 
$\frac{\omega_{\rm m}}{\sqrt{2}}$ &
$\frac{\omega}{\sqrt{2}} \sqrt{1-\frac{\omega^2}{\omega_{\rm m}^2}}\Theta(\omega_{\rm m}-\omega)$\\
\hline
\end{tabular}
\caption{\label{TAB:spectral_densities_phonon}
Phonon-type mappings for selected spectral densities according to Eqs.~(\ref{EQ:newphonon1}),~(\ref{EQ:newphonon2}), and~(\ref{EQ:phonon3}) 
with either a soft or a rigid ultraviolet cutoff.
All spectral densities are -- when analytically continued to the complete real axis -- odd functions of $\omega$.
All parameters have dimension of energy.
}
\end{table}

\subsubsection{Particle mapping}

In analogy with the phonon mapping case, we let the particle mapping be given by
\begin{align}\label{particlemap}
H &= H_S + S \sum_k h_k^* a_k^\dagger + S^\dagger \sum_k h_k a_k + \sum_k \omega_k a_k^\dagger a_k\nn
&= H_S + \lambda  S b^\dagger + \lambda S^\dagger b + \Omega b^\dagger b
+ b \sum_k H_k^* b_k^\dagger + b^\dagger \sum_k H_k b_k
+ \sum_k \Omega_k b_k^\dagger b_k\,.
\end{align}
Here, $S$ is again some dimensionless system operator, which need not necessarily be Hermitian, and it suffices to choose the Bogoliubov transform as unitary ($v_{kq}=0$), {see also Ref.~\cite{woods2014a} for examples.}
Then, from identifying $\lambda b = \sum_k h_k a_k$ and demanding bosonic commutation relations, we must have
\begin{align}\label{EQ:particle1a}
\lambda^2 &= \sum_k \abs{h_k}^2 = \frac{1}{2\pi} \int_0^\infty \Gamma^{(0)}(\omega) d\omega\,.
\end{align}
Second, we see that the relations $\sum_k h_k u_{k1} = \sum_k h_k^* u_{k1}^* = \lambda$
together with the unitarity relation $\sum_k \abs{u_{k1}}^2 = 1$ fix the first coefficients $u_{k1}=h_k^*/\lambda$, which when 
inserted into the energy of the reaction coordinate $\Omega= \sum_k \omega_k \abs{u_{k1}}^2$ eventually yields
\begin{align}\label{EQ:particle1b}
\Omega = \frac{1}{2\pi\lambda^2} \int_0^\infty \omega \Gamma^{(0)}(\omega) d\omega\,,
\end{align}
such that $\Omega$ also has dimensions of energy.

The Heisenberg equations (see Appendix~\ref{APP:particle_particle}) tell us that the following spectral density mapping relation should hold
\begin{align}\label{EQ:particle2}
\Gamma^{(1)}(\omega) &= \frac{4 \lambda^2 \Gamma^{(0)}(\omega)}
{\left[\frac{1}{\pi} {\cal P} \int_0^\infty \frac{\Gamma^{(0)}(\omega')}{\omega'-\omega} d\omega'\right]^2 + \left[\Gamma^{(0)}(\omega)\right]^2}\,.
\end{align}
Here, the difference to the phonon mapping is that $\Gamma^{(n)}(\omega)$ are not analytically continued to the complete real
axis, as $\omega>0$ is assumed throughout.
The convergence properties of related recursion relations have been discussed in great detail~\cite{woods2014a,woods2015a,woods2016a}.

As with the previous treatment, the structure of the Hamiltonian (bosonic tunneling) is similar before and after the
transformation, we just need to redefine the system and reservoir.
Therefore, it can also be applied recursively. 
This way, we can understand the reaction coordinate mapping as the sequential application of multiple Bogoliubov transformations.
%
%
In practise, one would truncate the resulting chain at some point, using a perturbative approach such as the master equation~\cite{iles_smith2014a,iles_smith2016a}.

Now, the spectrum of the supersystem Hamiltonian is bounded from below for any value of $\lambda$ when one can decompose it as
$H_S = H_S^0 + \frac{\lambda^2}{\Omega} S^\dagger S$, where $H_S^0$ is bounded from below. We finally note that by choosing $S=a+a^\dagger$ we can switch from a phonon-type representation to a particle-type representation.
%


\subsection{Fermionic particle mapping}

We can also investigate these problems for fermions, where the mapping reads
\begin{align}
H &= H_S + c \sum_k t_k^* c_k^\dagger - c^\dagger \sum_k t_k c_k + \sum_k \epsilon_k c_k^\dagger c_k\nn
&= H_S + \lambda  c d^\dagger - \lambda c^\dagger d + \varepsilon d^\dagger d
+ d \sum_k T_k^* d_k^\dagger - d^\dagger \sum_k T_k d_k
+ \sum_k \varepsilon_k d_k^\dagger d_k\,.
\end{align}
The difference is that the spectral densities can be defined also for negative frequencies, such that no analytic continuation is necessary.
Now, we obtain the reaction coordinate coupling and energy from integrals over the complete energies
\begin{align}\label{EQ:fermionic1}
\lambda^2 &= \frac{1}{2\pi} \int \Gamma^{(0)}(\omega) d\omega\,,\qquad
\varepsilon = \frac{1}{2\pi\lambda^2} \int \omega \Gamma^{(0)}(\omega) d\omega\,.
\end{align}
The Heisenberg equations (see Appendix~\ref{APP:fermion_fermion}) tell us that the following mapping relations should hold
\begin{align}\label{EQ:fermionic2}
\Gamma^{(1)}(\omega) &= \frac{4 \lambda^2 \Gamma^{(0)}(\omega)}
{\left[\frac{1}{\pi} {\cal P} \int \frac{\Gamma^{(0)}(\omega')}{\omega'-\omega} d\omega'\right]^2 + \left[\Gamma^{(0)}(\omega)\right]^2}\,,
\end{align}
where we again stress that the spectral density is defined also for negative energies.
When we consider the case that it strictly vanishes for negative energies, we recover the case of bosonic particle mappings from Eqs.~(\ref{EQ:particle1a}),~(\ref{EQ:particle1b}), and~(\ref{EQ:particle2}).

Table~\ref{TAB:spectral_densities_fermion} provides some examples of spectral densities and their mappings according to Eqs.~(\ref{EQ:fermionic1}) and~(\ref{EQ:fermionic2}). 
The functional form of the mapping implies that convergence of all integrals is ensured only for a hard cutoff.
In particular, the limiting case for particle mappings with a rigid cutoff is a semicircle, see the bottom right entry in 
Table~\ref{TAB:spectral_densities_fermion}, which we also observe numerically in Fig.~\ref{FIG:convergence} (right panel).
\begin{table}[h]
\begin{tabular}{|c|c|c|c|}
\hline
$\Gamma^{(0)}(\omega)$ & $\lambda^2$ & $\varepsilon$ & $\Gamma^{(1)}(\omega)$\\
\hline
\hline
$\Gamma \frac{\delta^2}{(\omega-\epsilon)^2+\delta^2}$ & 
$\frac{\Gamma \delta}{2}$ & 
$\epsilon$ & 
$2\delta$\\
\hline
$\Gamma \frac{\delta^4}{\left[(\omega-\epsilon)^2+\delta^2\right]^2}$ & $\frac{\Gamma \delta}{4}$ & $\epsilon$ & 
$\frac{\delta (2\delta)^2}{(\omega-\epsilon)^2+(2\delta)^2}$\\
\hline
$\Gamma e^{-\frac{(\omega-\epsilon)^2}{\delta^2}}$ & $\frac{\Gamma\delta}{2\sqrt{\pi}}$ & $\epsilon$ &
$\frac{2 \delta e^{+\frac{(\omega-\epsilon)^2}{\delta^2}}}{\sqrt{\pi}\left[1-{\rm erf}^2\left(\ii \frac{\omega-\epsilon}{\delta}\right)\right]}$
\\
\hline
$\Gamma \Theta(\omega,\epsilon-\delta,\epsilon+\delta)$ & $\frac{\Gamma\delta}{\pi}$ & $\epsilon$ &
$\frac{4\pi\delta}{\pi^2+4{\rm arctanh}^2\left(\frac{\epsilon-\omega}{\delta}\right)} \Theta(\omega,\epsilon-\delta,\epsilon+\delta)$\\
\hline
$\Gamma \left[1-\left(\frac{\omega}{\delta}-\frac{\epsilon}{\delta}\right)^2\right] \Theta(\omega,\epsilon-\delta,\epsilon+\delta)$ &
$\frac{2}{3} \frac{\Gamma \delta}{\pi}$ & $\epsilon$ & 
$\frac{8 \delta}{3\pi}\frac{\left[1-\frac{(\omega-\epsilon)^2}{\delta^2}\right]\Theta(\omega,\epsilon-\delta,\epsilon+\delta)}{\frac{4\left(\delta(\omega-\epsilon)-(\omega+\delta-\epsilon)(\omega-\delta-\epsilon)
{\rm arctanh}\left[\frac{\omega-\epsilon}{\delta}\right]\right)^2}{\pi^2\delta^4} + \left(1-\frac{(\omega-\epsilon)^2}{\delta^2}\right)^2}$
\\
\hline
$\Gamma \sqrt{1-\left(\frac{\omega}{\delta}-\frac{\epsilon}{\delta}\right)^2} \Theta(\omega,\epsilon-\delta,\epsilon+\delta)$ &
$\frac{\Gamma\delta}{4}$ &
$\epsilon$ &
$\delta \sqrt{1-\left(\frac{\omega}{\delta}-\frac{\epsilon}{\delta}\right)^2} \Theta(\omega,\epsilon-\delta,\epsilon+\delta)$\\
\hline
\end{tabular}
\caption{\label{TAB:spectral_densities_fermion}
Selected mappings for spectral densities according to Eqs.~(\ref{EQ:fermionic1}) and~(\ref{EQ:fermionic2}), using $\Theta(x,a,b)=\Theta(x-a)\Theta(b-x)$
and ${\rm erf}(z) = \frac{2}{\sqrt{\pi}} \int_0^z e^{-t^2} dt$.
{For the evaluation of the first row reaction coordinate energy $\varepsilon$, the principal value has to be taken, as the ordinary integral does not converge.}
As a rule of thumb, the width of the old spectral density becomes the coupling strength of the new spectral density, and only a rigid cutoff will
survive recursive transformations.
When the original spectral density strictly vanishes for negative $\omega$ (accessible with the last three rows for suitable parameters), 
we recover the bosonic particle mapping from
Eqs.~(\ref{EQ:particle1a}),~(\ref{EQ:particle1b}), and~(\ref{EQ:particle2}).
}
\end{table}
%


\subsection[Stationary State]{General Properties: Stationary state of the supersystem}

In the strong-coupling limit, we no longer expect the local Gibbs state $e^{-\beta H_S}/Z_S$ to be the stationary state of the system.
Rather, one might expect it to be given by the reduced density matrix of the total Gibbs state~\cite{gogolin2016a,perarnau_llobet2018a}
\begin{align}\label{EQ:redfullgibbs}
\bar\rho_S \approx {\rm Tr_B}\left\{\frac{e^{-\beta(H_S + H_B + H_I)}}{Z}\right\}\,,
\end{align}
which would only coincide with the system-local Gibbs state when $H_I\to 0$ (vanishingly weak coupling).
Since the reaction coordinate mappings allow for arbitrarily strong coupling between the original system and reservoir, 
we can test when the resulting stationary state in the supersystem is consistent with these expectations.

In particular, we assume here that the coupling between the supersystem and residual reservoir is small, such that we 
can apply the master equation formalism to the supersystem~\cite{iles_smith2014a,iles_smith2016a}
\begin{align}
H'_S = H_S + H_{RC} + H_I\,,
\end{align}
composed of system and reaction coordinate.
For the standard quantum-optical master equation (based in general on Born-Markov and secular approximations) it is known that for a single reservoir the stationary state
will approach the system-local Gibbs state~\cite{duemcke1979a,breuer2002}, now associated with the supersystem~\cite{iles_smith2014a}
\begin{align}\label{EQ:RCGibbs}
\bar\rho'_S = \frac{e^{-\beta H'_S}}{\ptrace{S,RC}{e^{-\beta H'_S}}}\,.
\end{align}
We define a Hamiltonian of mean force $H^*$ {-- {see also} the previous chapter --}
via the relation
\begin{align}
e^{-\beta H^*} = \frac{\traceB{e^{-\beta(H_S+H_I+H_B)}}}{\traceB{e^{-\beta H_B}}}\,.
\end{align}
It can be seen as an effective Hamiltonian for the system in the strong coupling limit.
In the weak-coupling limit ($H_I\to 0$) we get $H^* \to H_S$.
By construction, the Hamiltonian of mean force obeys
\begin{align}
e^{-\beta H^*} &= \frac{\ptrace{RC,B'}{e^{-\beta(H'_S+\tilde\lambda H'_I+H'_B)}}}{\ptrace{RC,B'}{e^{-\beta (H_{RC}+\tilde\lambda H'_I+H'_B)}}}
= \frac{\ptrace{RC}{e^{-\beta H'_S}}}{\ptrace{RC}{e^{-\beta H_{RC}}}} + \ord\{\tilde\lambda\}\,.
\end{align}
Here, $\tilde\lambda$ serves as a dimensionless bookkeeping parameter for the coupling between the reaction coordinate and the residual reservoir.
With Eq.~(\ref{EQ:RCGibbs}), this implies that the reduced steady state of the original system becomes
\begin{align}
\bar\rho_S &= \ptrace{RC}{\bar\rho'_S} = \frac{\ptrace{RC}{e^{-\beta H'_S}}}{\ptrace{S,RC}{e^{-\beta H'_S}}}
= \frac{e^{-\beta H^*} \ptrace{RC}{e^{-\beta H_{RC}}}}{\ptrace{S,RC}{e^{-\beta H'_S}}} + \ord\{\tilde\lambda\}
= \frac{e^{-\beta H^*}}{\ptrace{S}{e^{-\beta H^*}}} + \ord\{\tilde\lambda\}\,,
\end{align}
where the last equality follows directly from performing $\ptrace{S}{e^{-\beta H^*}}$.
That is, when the coupling $\tilde\lambda$ between the supersystem and the residual reservoir (i.e.~the transformed spectral density) is small, 
the approach recovers the reduced steady state~(\ref{EQ:redfullgibbs}) of the global Gibbs state~\cite{iles_smith2014a,strasberg2016a}.


\section{Applications to thermal machines}

Heat engines generate useful work by harnessing heat flow between hot and cold reservoirs. 
Usually, heat engine models are analysed under the simplifying assumption of negligibly weak interactions between the working system and the reservoirs. 
However, as argued earlier, for heat engines operating at the quantum scale such an approximation may not be well justified, since interaction energies potentially 
become comparable with system and reservoir self-energy scales. 
The treatment becomes even more challenging when one allows for driven heat engines, where the coupling strength between the system and reservoir is
modified periodically.
In such cases, even the correct partition of the time-dependent system-reservoir driving into heat and work contributions is generally a challenging task, where the
reaction coordinate treatment is indeed helpful~\cite{restrepo2018a}.
For conceptual simplicity, however, we shall put the simultaneous treatment of driving and dissipation aside and therefore review two types of heat engine model in this section -- discrete stroke and continuous -- that have recently been analysed 
beyond weak system-reservoir coupling by employing the reaction coordinate formalism outlined above.


\subsection[Discrete stroke engines]{Discrete stroke engines: the Otto cycle}

Discrete stroke heat engines operate in closed cycles that are divided into individual sections (strokes) in which particular operations such as heat exchange, expansion, compression, or combinations, take place. The working system returns to its original state at the end of the cycle and in order to produce a finite power output, each stroke must be performed within a finite time. Nevertheless, the study of infinite time cycles (which produce work but zero power output) has been crucial in identifying fundamental thermodynamic bounds on heat engine performance, and indeed led to Carnot's principle. Furthermore, the identification of a closed cycle is made easier in the infinite time limit due to the knowledge of the state (whether within the weak-coupling or reaction coordinate approaches) after equilibration between the system and the hot or cold reservoir. Thus, zero power cycles provide a natural setting in which to begin exploring extensions of heat engine models beyond the weak system-reservoir coupling regime.

Here we shall review the reaction coordinate analysis of a discrete stroke quantum Otto cycle beyond weak reservoir coupling as presented in Ref.~\cite{newman2017a}. This cycle is a quantum analogy of a four-stroke internal combustion engine model. It has the advantage of separating strokes in which energy is exchanged between the system and the reservoirs from those in which work is either extracted from or done on the system with the reservoirs uncoupled. In this way, we seek to avoid the complications of defining work and heat at strong-coupling for strokes in which both may be important, as encountered for example within the Carnot cycle.

\begin{figure}
\includegraphics[width=0.48\textwidth]{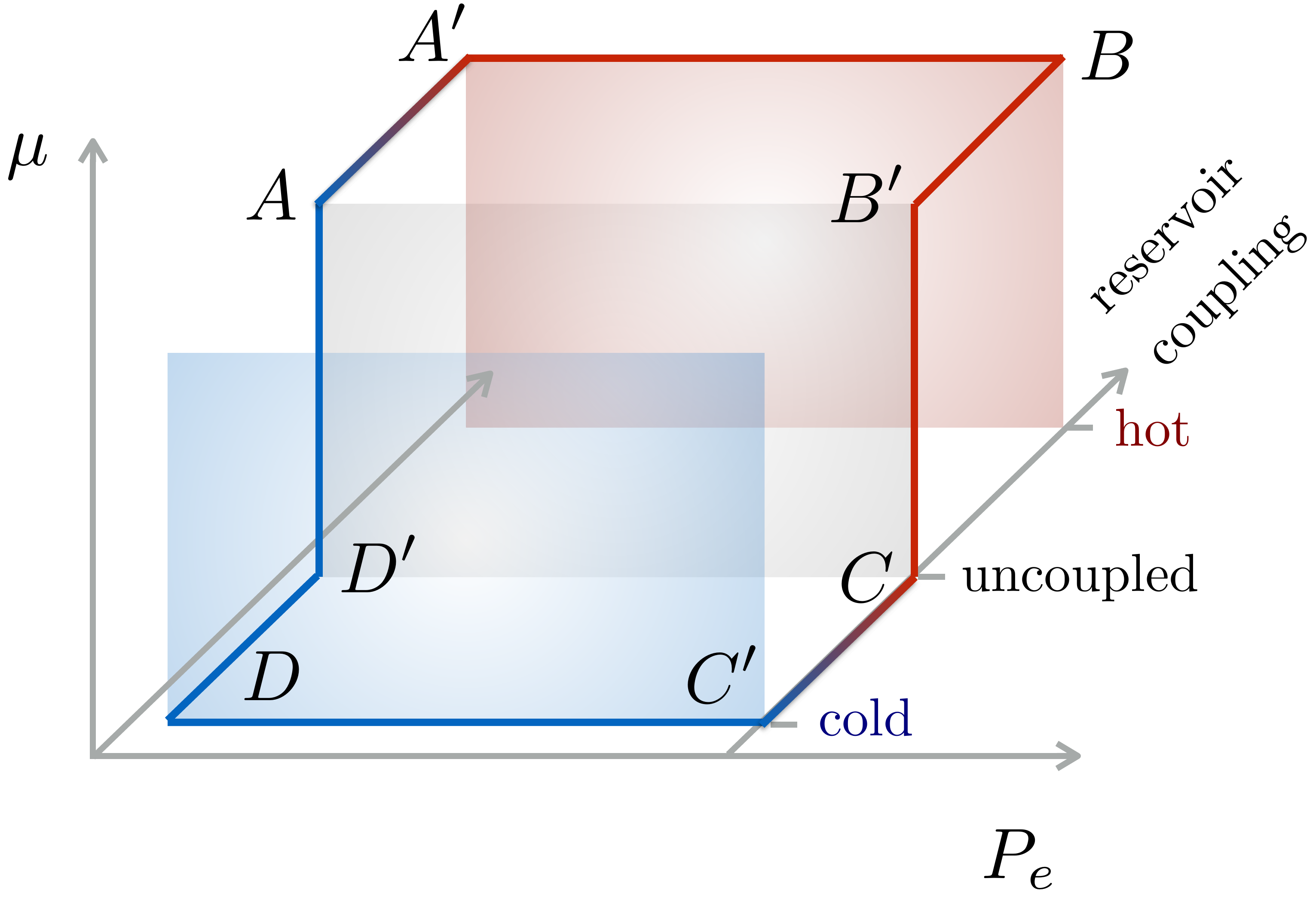}
\caption{Schematic of a quantum Otto cycle for a two-level system with splitting $\mu$ (vertical axis) and excited state population  $P_e$ (horizontal axis), which may be coupled to and decoupled from hot and cold reservoirs (red and blue shaded areas, respectively). The depicted cycle strokes are: isochoric equilibration $A' \rightarrow B$, decoupling from the hot reservoir $B\rightarrow B'$, isentropic expansion $B' \rightarrow C$, coupling to the cold reservoir $C\rightarrow C'$, isochoric equilibration $C' \rightarrow D$, decoupling from the cold reservoir $D\rightarrow D'$, isentropic compression $D' \rightarrow A$, and coupling to the hot reservoir $A\rightarrow A'$. Reprinted with permission from D.~Newman, F.~Mintert, and A.~Nazir, Physical Review E 95, 032139 (2017), \url{https://dx.doi.org/10.1103/PhysRevE.95.032139}. Copyright (2017) by the American Physical Society. 
\label{ottofig}}
\end{figure}

The strokes of our quantum Otto cycle are depicted schematically in Fig.~{\ref{ottofig}}. We consider a two-level working system with ground state $|g\rangle$ and excited state $|e\rangle$, split by an energy $\mu$. The system may be coupled to and decoupled from two reservoirs, one hot and one cold. 
In the standard approach assuming negligibly weak system-reservoir interactions these coupling and decoupling steps make no energetic contribution to the cycle. This can be seen, for example, if we consider the system to couple linearly to bosonic reservoirs that remain in thermal equilibrium (factorised from the system state) throughout the cycle, in which case the trace of the interaction Hamiltonian evaluates to zero. Under these conditions, the Otto cycle consists of four strokes, two isentropes in which the system Hamiltonian is varied in time in the absence of any reservoir coupling, and two isochores in which the system equilibrates with either the hot or cold reservoir without any variations in the system Hamiltonian. However, for the case of finite reservoir coupling that is of interest here, there is no reason to believe that the coupling and decoupling steps can be neglected. We must therefore enlarge the cycle to include these contributions as well, leading to the primed points in Fig.~\ref{ottofig}.

Let us now analyse the cycle in more detail, contrasting the reaction coordinate and weakly-interacting approaches. Starting, arbitrarily, at point $A'$ in Fig.~\ref{ottofig}, the system and hot reservoir have just been coupled. They are allowed to equilibrate along the subsequent stroke, known as the hot isochore, which means that within the weak-coupling limit the state of the system plus hot reservoir is given at point $B$ by 
\begin{align}\label{thermalfactor}
\bar{\rho}_S^0\otimes\rho_h=\frac{e^{-\beta_hH_S}}{{\rm Tr_S}\left\{e^{-\beta_hH_S}\right\}}\otimes\frac{e^{-\beta_hH_B}}{{\rm Tr_S}\left\{e^{-\beta_hH_B}\right\}}.
\end{align}
Here, $\beta_h=1/T_h$ is the inverse temperature of the hot reservoir with internal Hamiltonian $H_B$. For weak coupling, the system simply thermalises along the stroke with respect to 
its internal Hamiltonian $H_S$. The system-reservoir coupling strength thus plays no role in the infinite time weak-coupling limit. In contrast, within the reaction coordinate formalism the state at the end of the stroke is given by
\begin{align}\label{isochoreRC}
\bar{\rho}'_S\otimes\tilde{\rho}_h=\bar{\rho}'_S\otimes\frac{e^{-\beta_h\tilde{H}_B}}{{\rm Tr_S}\left\{e^{-\beta_h\tilde{H}_B}\right\}},
\end{align}
where $\bar{\rho}'_S$ is now a Gibbs (thermal) state of the supersystem comprised of both the original two-level system and the reaction coordinate, as defined in Eq.~(\ref{EQ:RCGibbs}) with $\beta\rightarrow\beta_h$. This encodes correlations due to finite interactions between the system and reservoir, and thus has a natural dependence on the system-reservoir coupling strength as well as the reservoir temperature. Note that the factorisation in Eq.~(\ref{isochoreRC}) is made only with respect to the mapped residual bath with internal Hamiltonian $\tilde{H}_B$, given for example by the final terms in Eqs.~(\ref{EQ:phononmap}) and (\ref{particlemap}). It is not, therefore, equivalent to a weak-coupling approximation between the system and the full reservoir as in Eq.~(\ref{thermalfactor}). Accordingly, numerical benchmarking of the reaction coordinate method has shown it to be accurate over a very wide range of system-environment coupling strengths~\cite{iles_smith2014a,iles_smith2016a}.

The quantity of interest for analysing the cycle performance is the energy expectation value at the end of each stroke. In the weak coupling limit only changes to the two-level system energy are tracked, and so we consider 
\begin{align}
\langle H\rangle_{\rm weak}={\rm Tr}\{H_S\bar{\rho}_S^0\}.
\end{align}
In the reaction coordinate approach, on the other hand, changes to both the system and reaction coordinate are monitored, and so we have 
\begin{align}
\langle H\rangle={\rm Tr}\{H_S'\bar{\rho}'_S\},
\end{align}
which includes additional contributions from the reaction coordinate and system-reaction coordinate Hamiltonians through $H_S'$, as well as correlation effects through $\bar{\rho}'_S$.\footnote{Note that there is a subtlety in the strong-coupling cycle. When coupled, the interaction between the system and the reservoir pushes the latter out of thermal equilibrium. We assume that once the system and reservoir are decoupled at the end of the stroke, the reservoir rapidly relaxes back to equilibrium. Hence, when the system comes to be coupled to the reservoir again on the next cycle, the reservoir is thermal once more. The re-thermalisation of the reservoirs entails accounting for some extra energetic contributions around the cycle, as described in detail in~\cite{newman2017a}.}

\begin{figure}
\includegraphics[width=0.48\textwidth]{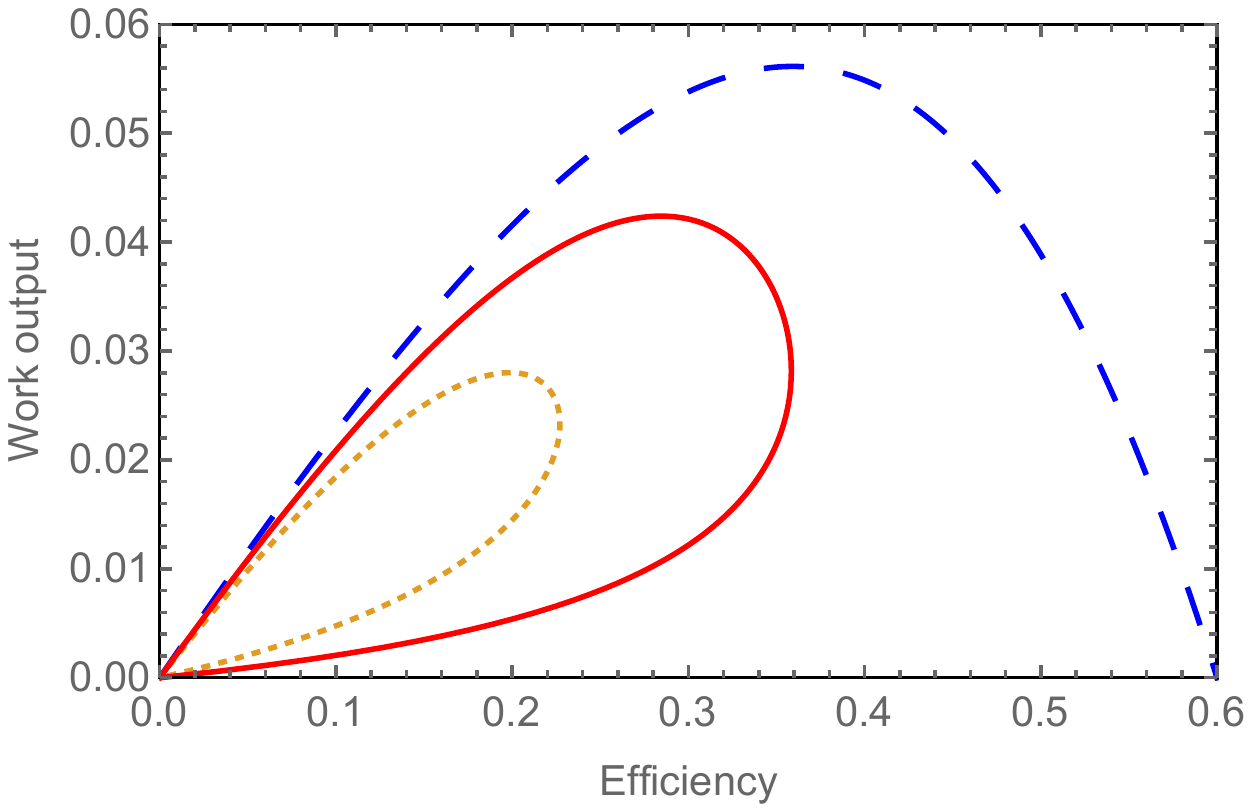}
\caption{Example parametric plots of Otto cycle work output and efficiency. Each curve is generated by varying the difference in splittings $\mu$ between the start and the end of the isentropic strokes. The blue dashed curve shows the weak-coupling treatment. The red solid and orange dashed curves show the reaction coordinate treatment with adiabatic and instantaneous decoupling of the reservoirs, respectively. Reprinted with permission from D.~Newman, F.~Mintert, and A.~Nazir, Physical Review E 95, 032139 (2017), \url{https://dx.doi.org/10.1103/PhysRevE.95.032139}. Copyright (2017) by the American Physical Society. 
\label{parametric}}
\end{figure}

At point $B$ the interaction between the system and hot reservoir is now turned off to reach point $B'$, which we must explicitly account for within the reaction coordinate analysis. For simplicity we shall assume this happens suddenly, such that the full state does not change, and hence define a work cost associated with decoupling of
\begin{align}
{\rm Tr}\{(H_S+H_{RC}-H_S')\bar{\rho}'_S\}=-{\rm Tr}\{H_{I}\bar{\rho}'_S\}.
\end{align}
This cost impacts adversely on the total work output of the cycle. In Ref.~\cite{newman2017a} it is shown that part (though not all) of the work cost can be mitigated by decoupling the system and reservoir slowly (i.e.~in the adiabatic limit), see also Fig.~\ref{parametric}.

From point $B'$ to $C$ the system Hamiltonian is changed such that the splitting is reduced ($\mu_{\rm B'}>\mu_{\rm C}$) in the absence of any reservoir coupling, with the stroke thus being termed isentropic expansion. For changes that are slow enough to justify use of the quantum adiabatic theorem, the average system energy along the stroke is given simply by
\begin{align}
\langle H_S(t)\rangle=\frac{\mu(t)}{\mu(0)}\langle H_S(0)\rangle,
\end{align}
where $t=0$ refers to the start of the stroke, $\langle H_S(0)\rangle={\rm Tr}\{H_S(0)\bar{\rho}_S^0(0)\}$ in the weak-coupling case, and $\langle H_S(0)\rangle={\rm Tr}\{H_S(0)\bar{\rho}'_S(0)\}$ in the reaction coordinate treatment. Thus, the work output of the stroke becomes
\begin{align}
W_{\rm stroke}=\left(\frac{\mu_{\rm C}}{\mu_{\rm B'}}-1\right)\langle H_S(0)\rangle,
\end{align}
which is negative by convention. 

The coupling to the cold reservoir is now switched on ($C\rightarrow C'$). We assume the reservoirs relax back to thermal equilibrium when decoupled from the system, and so there is no work contribution associated with this step in either treatment as the trace of the interaction Hamiltonian is then zero. The system and cold reservoir now equilibrate along the subsequent stroke ($C'\rightarrow D$), which we may analyse in the same way as the hot isochore. They are then decoupled ($D\rightarrow D'$) once more incurring a work cost in the reaction coordinate treatment. Subsequently, we do work on the system during an isentropic compression ($D'\rightarrow A$). Here, the system splitting is adiabatically increased back to its earlier value, such that $\mu_A=\mu_{B'}$. Finally, the system and hot reservoir are coupled ($A\rightarrow A'$, no work cost) and the cycle is complete.

In Fig.~{\ref{parametric}} we show some example parametric plots of the efficiency and work output of the Otto cycle as treated within both the weak-coupling and reaction coordinate frameworks. To generate these curves the difference in splittings between the start and end of the isentropic stokes is varied. In the weak-coupling case, the net work output is simply the difference in work along the two isentropic strokes, whereas in the reaction coordinate treatment we must include the decoupling costs as well. In both cases, we define the efficiency in the standard way as
\begin{align}
\eta=\frac{W}{Q_{A'B}},
\end{align}
where $W$ is the cycle net work output and $Q_{A'B}$ is the energy change along the hot isochore. For the weak-coupling case this can be expressed in the simple form $\eta_{\rm weak}=1-\mu_C/\mu_{B'}$, which depends only on the ratio of the two-level splittings. We can see from Fig.~{\ref{parametric}} that the weak-coupling efficiency reaches a maximum at the point at which the work output vanishes. Here, the ratio of splittings becomes equal to the ratio of cold to hot reservoir temperatures, $T_c/T_h$, and so the weak-coupling efficiency reaches the Carnot bound. The behaviour of the cycle is qualitatively different, and inferior, in the reaction coordinate case. The maximum efficiency occurs at finite work output, though takes values well below the Carnot bound, and the efficiency also falls to zero as the work output vanishes at larger ratios of the two-level splittings. The decoupling cost contributions are the primary cause for the reduced engine performance at strong-coupling, with a small reduction in energy absorbed along the hot isochore insufficient to overcome their detrimental effect to the efficiency~\cite{newman2017a}. Finally, we note that an adiabatic decoupling protocol can improve both efficiency and work output (though not up to the idealised weak-coupling limit), and should thus be an important consideration in optimising the performance of nanoscale (quantum) engine cycles where the presence of non-negligible reservoir couplings is expected~\cite{lehur2012a,goldstein2013a,peropadre2013a,nazir2016a}.


\subsection[Continuous engines]{Continuously operating thermo-machines}

For continuously operating heat engines (or refrigerators), the system of interest is coupled to multiple reservoirs that
are held at different local thermal equilibrium states throughout~\cite{kosloff2014a}.
It is then possible to use, for example, a thermal gradient between the reservoirs to extract (chemical) work by transporting electrons against a bias (heat engine)
or to cool the coldest reservoir by investing work (chemical work or the energy provided by a so-called work reservoir).
In this section, we shall exemplarily benchmark the reaction coordinate treatment of an exactly-solvable two-terminal model. 

The single-electron transistor (SET) with Hamiltonian
\begin{align}
H = \epsilon d^\dagger d  + \sum_{k\alpha} \left[ t_{k\alpha} d c_{k\alpha}^\dagger + {\rm h.c.}\right] + \sum_{k\alpha} \epsilon_{k\alpha} c_{k\alpha}^\dagger c_{k\alpha}
\end{align} 
describes a single quantum dot $d$ with on-site energy $\epsilon$ that is coupled via tunneling amplitudes $t_{k\alpha}$ to two fermionic leads $\alpha\in\{L,R\}$.
Letting the leads become continuous, we introduce the original lead spectral densities $\Gamma_\alpha^{(0)}(\omega) = 2\pi \sum_k \abs{t_{k\alpha}}^2 \delta(\omega-\epsilon_{k\alpha})$.
The model can be analyzed as a heat engine with a perturbative treatment of the $t_{k\alpha}$~\cite{esposito2009b}.
However,  an exact solution can also be derived~\cite{haug2008} and analyzed from a thermodynamic viewpoint~\cite{topp2015a}.
We consider reservoirs described by temperatures $\beta_\alpha$ and chemical potentials $\mu_\alpha$ and use conservation of energy and matter currents at steady state throughout.

One observable of interest is then the chemical work rate extracted from the system
\begin{align}
P = -(\mu_L - \mu_R) I_M\,,
\end{align}
where $I_M$ denotes the electronic matter current counting positive from left to right.
For $P>0$, this process can be interpreted as electric power used to transport electrons against a bias voltage $V=\mu_L-\mu_R$.
Furthermore, we define the stationary heat currents entering the system as
\begin{align}
\dot{Q}_L = I_E - \mu_L I_M\,,\qquad
\dot{Q}_R = -(I_E - \mu_R I_M)\,,
\end{align}
where $I_E$ denotes the energy current counting positive from left to right.
Without loss of generality, we consider setups where $\mu_L>\mu_R$ and $\beta_L > \beta_R$ ($T_L<T_R$).
Then, the efficiency of generating electric power $P>0$ from the heat coming from the hot reservoir $\dot{Q}_R>0$ becomes
\begin{align}\label{EQ:efficiency}
\eta = \frac{P \Theta(P)}{\dot{Q}_R} \le 1 - \frac{T_L}{T_R} = \eta_{\rm Ca}\,,
\end{align}
where $\Theta(P)$ denotes the Heaviside-$\Theta$ function.
Here, the upper bound by Carnot efficiency follows from the positivity of the entropy production 
rate, which at steady state reduces to $\dot{S}_\ii = -\beta_L \dot{Q}_L - \beta_R \dot{Q}_R\ge 0$~\cite{topp2015a}.
With the same argument, the coefficient of performance for cooling the cold reservoir $\dot{Q}_L>0$ by investing chemical work $P<0$,
\begin{align}\label{EQ:cop}
{\rm COP} = \frac{\dot{Q}_L \Theta(\dot{Q}_L)}{-P} \le \frac{T_L}{T_R-T_L} = {\rm COP}_{\rm Ca},
\end{align}
must also obey a Carnot bound.

\begin{figure}
\centering 	
\setlength{\unitlength}{.1\linewidth}
\begin{picture}(10,5)
	\put(0.,0.){\includegraphics[width=.484\linewidth]{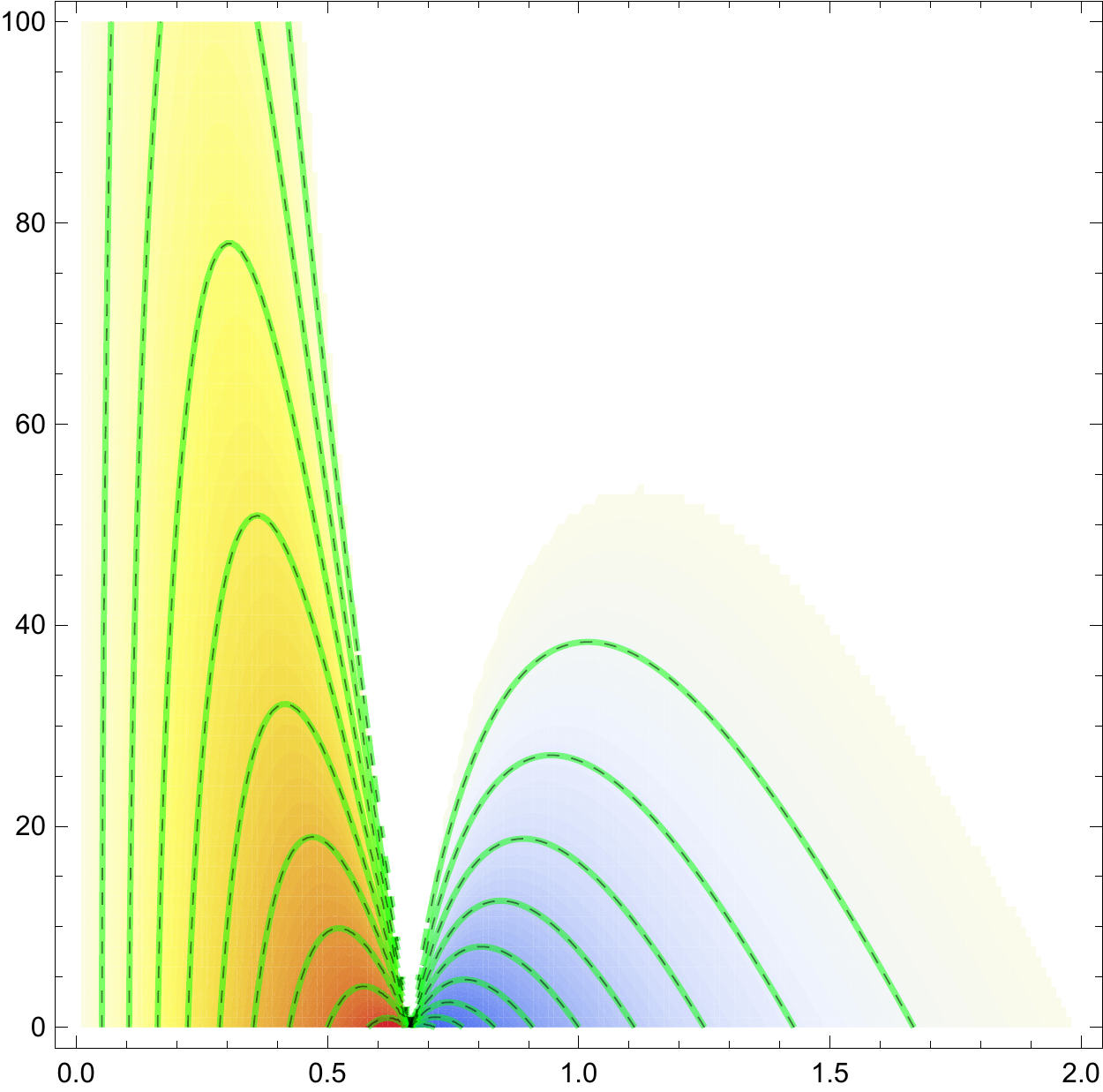}}
	\put(1.5,-.2){bias voltage $V/\epsilon$}
	\put(-.1,1.5){\rotatebox{90}{coupling strength $\Gamma/\epsilon$}}
	\put(4.25,1.5){\includegraphics[width=0.05\linewidth]{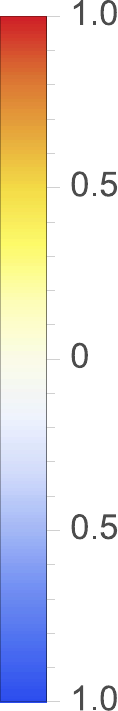}}
	\put(5.,0.){\includegraphics[width=.5\linewidth]{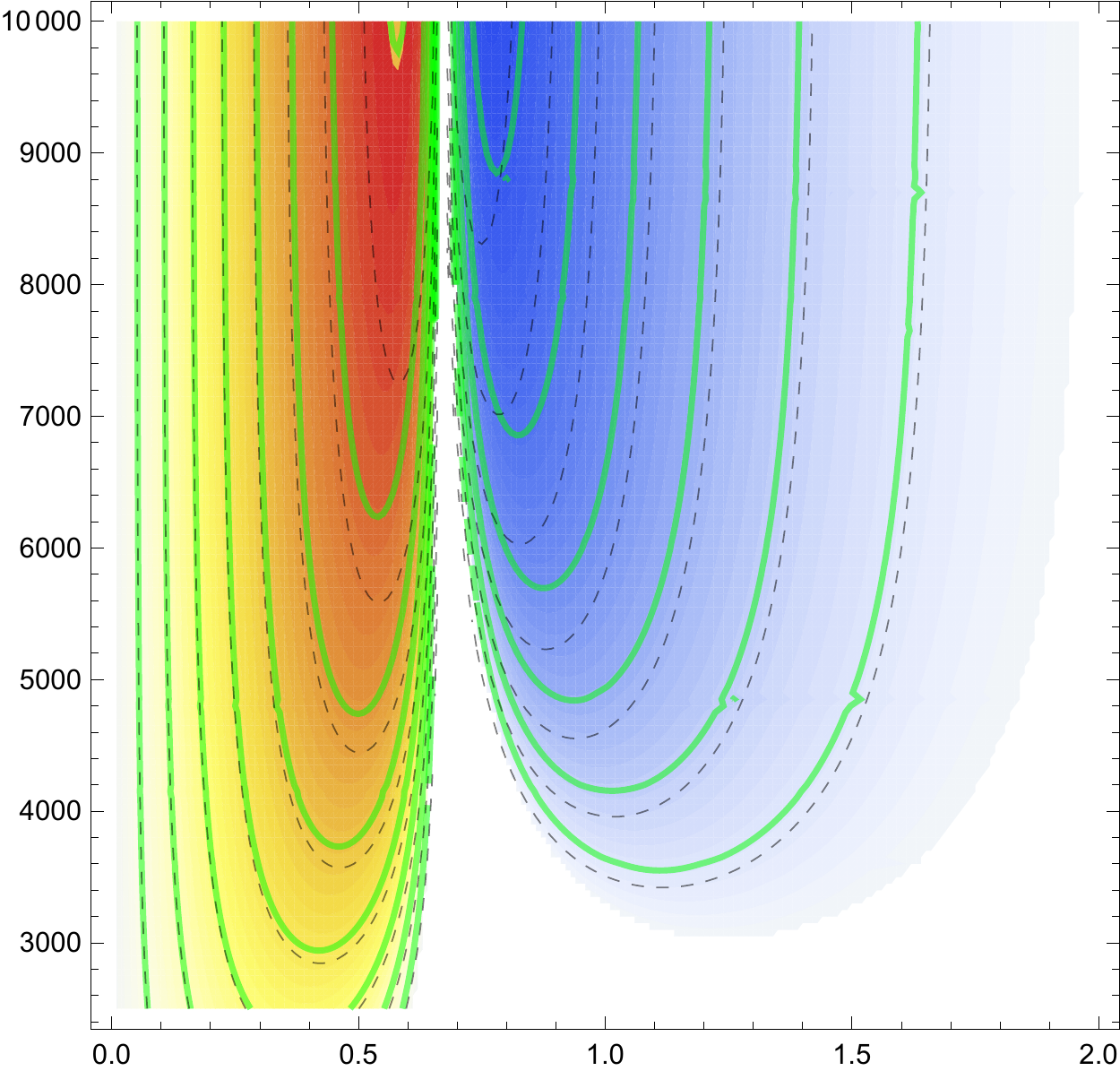}}
	\put(6.5,-.2){bias voltage $V/\epsilon$}
	\put(4.9,1.5){\rotatebox{90}{coupling strength $\Gamma/\epsilon$}}
\end{picture}
\caption{\label{FIG:coupling}
{\bf Left:}
Density plot of the heat engine (red colours) and cooling (blue colours) efficiencies in units of their maximum Carnot value 
versus dimensionless bias voltage $V/\epsilon$ 
(horizontal axis) and coupling strength $\Gamma/\epsilon$ (vertical axis).
Colours and solid green contour lines (in steps of $.1$) correspond to the exact solution of the SET.
They agree perfectly with the thin dashed contour lines, which have been calculated using a Born-Markov master 
equation treatment of the triple dot (reaction coordinate) supersystem.
At vanishing coupling strength (bottom), the maximum Carnot values are reached, and we see a direct transition from heat engine to refrigerator operational modes.
For stronger couplings, a gap between these modes opens, performances decrease, and while for strong couplings the cooling mode vanishes completely, 
the device may still act as a heat engine -- albeit at reduced efficiency.
Other parameters: $\Gamma_L=\Gamma_R=\Gamma$, $\delta_L=\delta_R=0.01\epsilon$, $\epsilon_L=\epsilon_R=\epsilon$, $\mu_L=-\mu_R=V/2$, $\beta_R\epsilon=1$, $\beta_L\epsilon=2$.
{\bf Right:}
Continuation of the left panel towards the ultrastrong coupling regime (with otherwise identical parameters).
Heat engine efficiency increases again and also cooling function is revived.
}
\end{figure}
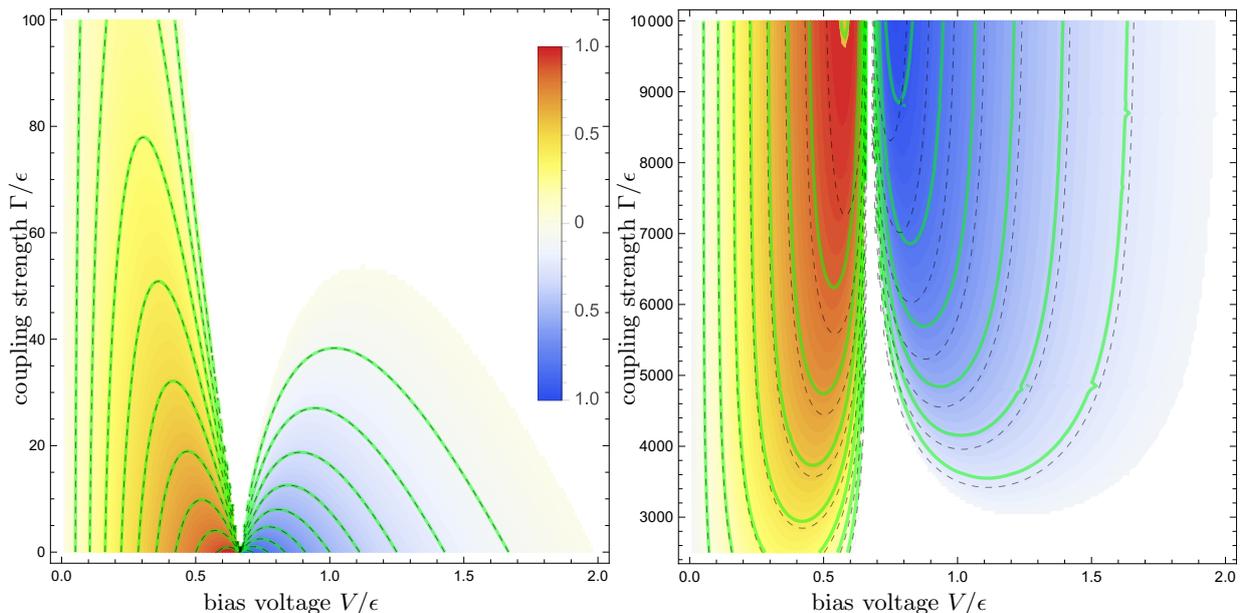

We can exactly evaluate energy and matter currents for the SET model using for example nonequilibrium Green's function
techniques~\cite{haug2008,bruch2016a} or other approaches~\cite{topp2015a}, which eventually allows for an exact evaluation of heat 
engine efficiency~(\ref{EQ:efficiency}) and COP~(\ref{EQ:cop}) for arbitrary system-reservoir coupling strengths.
Alternatively, we can apply the fermionic reaction coordinate mapping~\cite{strasberg2018a}, which 
-- when we use a separate reaction coordinate for every reservoir -- transforms the SET to a triple quantum dot
that is tunnel-coupled to two residual leads~\cite{schaller2018a}
\begin{align}\label{EQ:ham_tqd}
H &= \epsilon d^\dagger d + \sum_\alpha \lambda_\alpha (d d_\alpha^\dagger + d_\alpha d^\dagger) + \sum_\alpha \epsilon_\alpha d_\alpha^\dagger d_\alpha
+\sum_{k\alpha} \left[T_{k\alpha} d_\alpha d_{k\alpha}^\dagger + {\rm h.c.}\right] + \sum_{k\alpha} \tilde{\epsilon}_{k\alpha} d_{k\alpha}^\dagger d_{k\alpha}\,,
\end{align}
where we have reaction coordinate couplings $\lambda_\alpha$ and energies $\epsilon_\alpha$ as well as a new spectral density 
$\Gamma^{(1)}(\omega) = 2\pi \sum_k \abs{T_{k\alpha}}^2 \delta(\omega-\tilde{\epsilon}_{k\alpha})$.
Specifically, using the Lorentzian spectral density from the first row of Table~\ref{TAB:spectral_densities_fermion} centred around $\epsilon_\alpha$ with coupling strengths $\Gamma_\alpha$ and width $\delta_\alpha$,
we see that the resulting triple quantum dot is tunnel-coupled to two residual reservoirs with a flat spectral density, to which a Markovian treatment of the supersystem (first three terms
in the above equation) should apply when $\beta_\alpha\delta_\alpha$ is small.
For this supersystem, we can set up the (non-secular) master equation (compare with Ref.~\cite{schaller2018a} in absence of feedback control operations)
and compute energy and matter currents entering the supersystem from the residual reservoirs.
At steady state, we can identify these with the original currents defined above (the reaction coordinates can only host finite charge/energy) and therefore
likewise evaluate heat engine efficiency and COP within the reaction coordinate formalism.
The result of this procedure is depicted in Fig.~\ref{FIG:coupling} (left panel).
There we see that at vanishing coupling, where the conventional single dot master equation approach to the SET applies~\cite{esposito2009b}, maximal efficiencies are actually 
reached (albeit at zero power or cooling current, respectively), and the transition between heat engine and cooling operational modes happens directly.
This can be understood since in the limit of vanishing coupling, the SET obeys the so-called {\em tight-coupling} 
condition $I_E = \epsilon I_M$.
For finite coupling strengths $\Gamma$, a gap between these modes opens, which is also observed in other models beyond the weak-coupling limit~\cite{restrepo2018a}. 
When we further increase the coupling strength, the cooling function is no longer attainable, and also the efficiency of the heat engine decreases.
Most importantly, we see that the reaction coordinate treatment (dashed contours), reproduces the exact solution (colours and solid green contours) well, 
which is attributable to the fact that we choose initially highly
peaked spectral densities, such that the residual coupling $\beta_\alpha\delta_\alpha$ is very small and the reaction coordinate treatment is valid.

One might now be tempted to think that larger coupling strengths are always detrimental to the perfomance of thermoelectric devices, compare
Ref.~\cite{perarnau_llobet2018a} or the discussion in the previous subsection.
With the exact solution at hand, there is little intuitive evidence for finding other interesting parameter regimes.
However, going further towards ultra-strong coupling reveals that there is a regime where the heat engine efficiency increases again, and
even the cooling operational mode is revived, as is illustrated in Fig.~\ref{FIG:coupling} (right panel). 
From comparing the contours we can see that the reaction coordinate treatment (dashed) correctly predicts the 
revival of the cooling mode and the strengthening of the heat engine efficiency in this limit.
{We note that as the tunnel amplitudes scale only with the root of the coupling strength $T_{k\alpha} \le \sqrt{\Gamma_{\rm max} \delta/2} \approx 32 \epsilon$ for the 
parameters in Fig.~\ref{FIG:coupling}, this regime is actually not unrealistic, as has been experimentally demonstrated in various quantum dot systems~\cite{baines2012a,hensgens2017a,bayer2017a}.
From the experimental side, the challenge is rather the maintenance of a thermal gradient.
}

\begin{figure}[t]
\includegraphics[width=0.3\textwidth,clip=true]{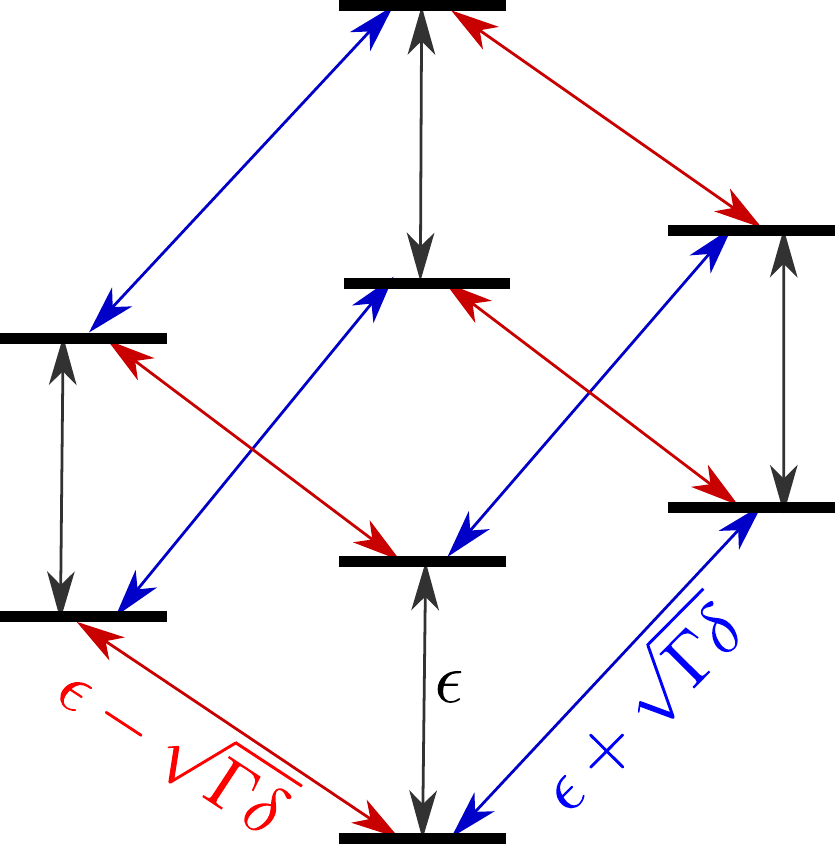}
\caption{\label{FIG:rategraph_bms_nfb}
Sketch of the supersystem spectrum consisting of the vacuum state (bottom), three singly charged states, three doubly occupied states, and the
fully charged state (top).
For $\delta_L=\delta_R=\delta$, $\Gamma_L=\Gamma_R=\Gamma$, and $\epsilon_L=\epsilon_R=\epsilon$, the allowed transitions (non-vanishing matrix elements of 
operators $d_{L/R}^{(\dagger)}$) only admit three transition frequencies:
$\Delta E = \epsilon$ (black arrows), $\Delta E = \epsilon - \sqrt{\Gamma \delta}$ (red arrows), and $\Delta E = \epsilon+\sqrt{\Gamma\delta}$
(blue arrows).
}
\end{figure}

\begin{figure}[ht]
\begin{tabular}{cc}
\includegraphics[width=0.48\textwidth]{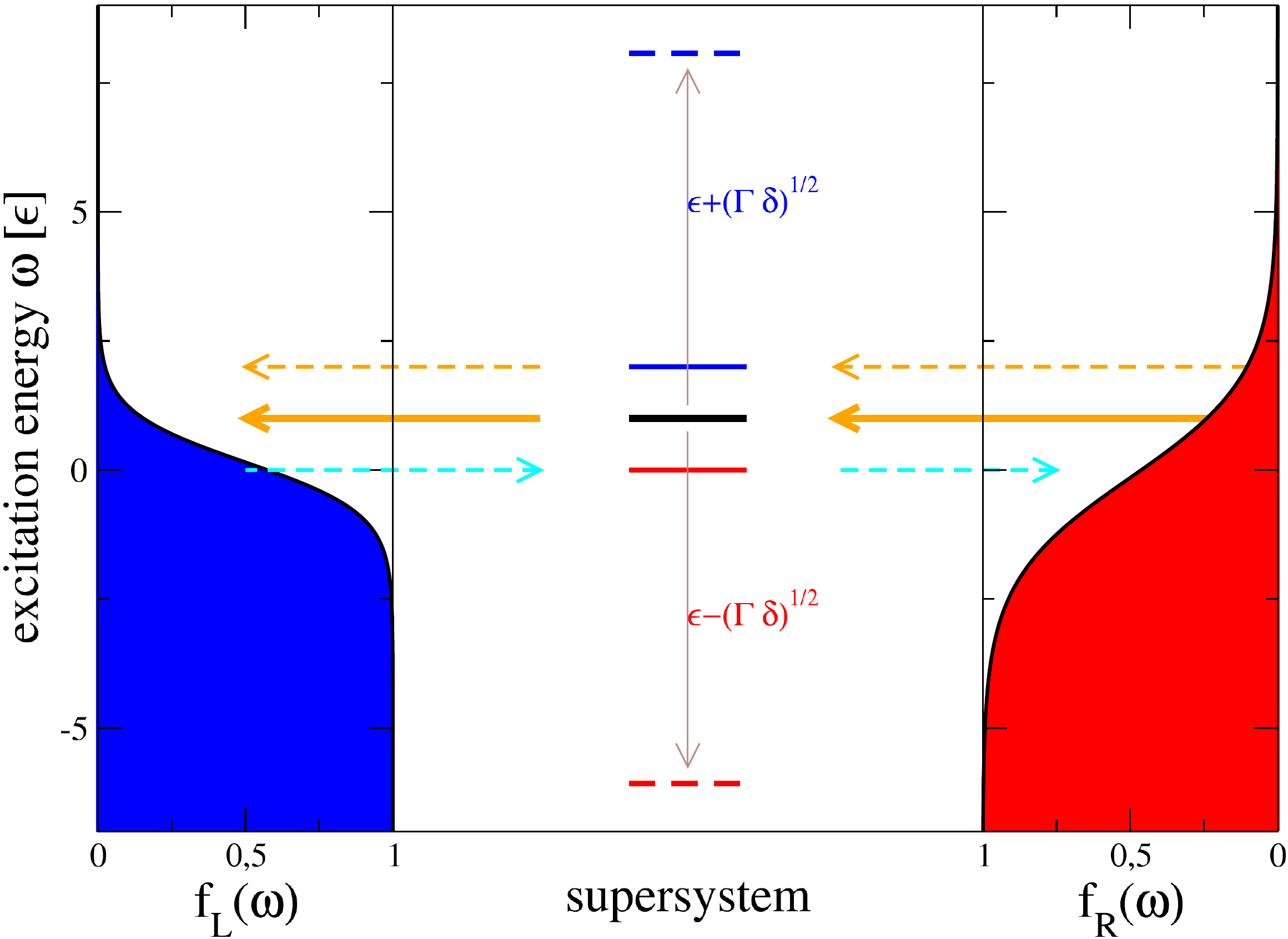} &
\includegraphics[width=0.48\textwidth]{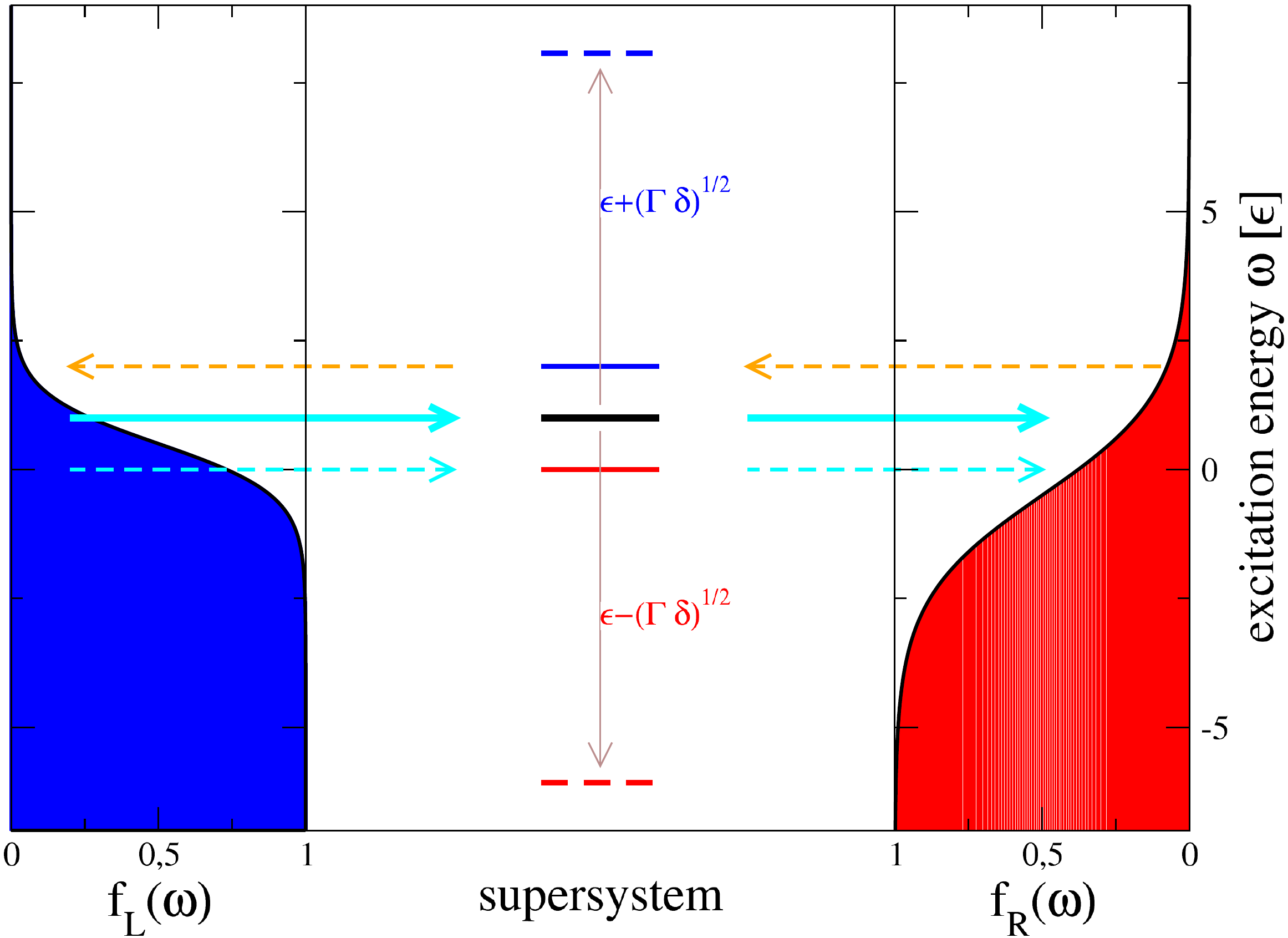}
\end{tabular}
\caption{\label{FIG:transportwindow}
Illustration of the supersystem transport window in the heat engine regime $V/\epsilon=0.3$ (left) and in the cooling regime $V/\epsilon=1.0$ (right).
Coloured regions left and right denote actual Fermi functions at the parameters used in the calculations.
Levels in the centre represent excitation energies of the supersystem depending on the coupling strength, and the current through
each transition is directed from the lead with larger Fermi function to the lead with the lower Fermi function.
In the heat engine regime (left), transport of electrons against the bias is induced by a thermal gradient (solid orange arrows). 
In the cooling regime (right), transport is driven by the voltage gradient and cools the cold (left) reservoir (solid blue arrows).
For small couplings, the three transition frequencies essentially merge into one at $\epsilon$ (black) and the reservoirs cannot resolve between them. We 
then approximately have tight coupling $I_E \approx \epsilon I_M$ and the naive (weak-coupling) single dot master equation applies.
For stronger couplings, two transition energies~(\ref{EQ:transition_energies}) split (solid red and blue for $\Gamma =100 \epsilon$), 
destroying tight coupling $I_E \neq \epsilon I_M$, reducing efficiencies, and even inhibiting the cooling function as one level supports a current with opposite sign (dashed arrows).
For ultrastrong couplings, the two variable transition frequencies have left the transport window (dashed red and blue for $\Gamma = 5000 \epsilon$) and thus no longer participate in transport, such that 
tight coupling $I_E \approx \epsilon I_M$ is restored.
Other parameters have been chosen as in Fig.~\ref{FIG:coupling}.
}
\end{figure}

Within the reaction coordinate picture, this worsening and re-strengthening of performance can be understood with a simple transport spectroscopy interpretation. 
It is rather straightforward to diagonalize the supersystem Hamiltonian in the first three terms of Eq.~(\ref{EQ:ham_tqd}).
Between the 8 energy eigenstates of the supersystem, not all transitions are allowed in the sequential tunneling regime, see Fig.~\ref{FIG:rategraph_bms_nfb}.
Adopting equal coupling strengths $\Gamma_L=\Gamma_R\equiv\Gamma$ and widths $\delta_L=\delta_R\equiv\delta$ for both reservoirs, and matching the 
maximum tunneling rate with that of the central dot $\epsilon_L=\epsilon_R=\epsilon$, the allowed transition frequencies simplify to
\begin{align}\label{EQ:transition_energies}
\Delta E \in\left\{ \epsilon, \epsilon-\sqrt{\Gamma \delta}, \epsilon+\sqrt{\Gamma\delta}\right\}\,.
\end{align}
With modifying the coupling strength $\Gamma$, we therefore change the transition energies as well.
For vanishingly small coupling strengths $\Gamma$, the splitting is not resolved by the rather large reservoir temperatures as 
$\beta_\alpha \sqrt{\Gamma\delta} \ll 1$, and the only visible transition frequency is $\epsilon$.
{As mentioned, in this regime the setup approximately obeys the tight-coupling property $I_E \approx \epsilon I_M$, 
and maximum efficiencies are reached (albeit at vanishing power)~\cite{vandenbroeck2005a,gomez_marin2006a,sheng2013a}.
Here,} even the naive (weak-coupling) SET master equation treatment is valid.
When we increase the coupling strength, we leave the tight-coupling regime, i.e.~energy and matter currents are no longer proportional to each other, which is known to decrease efficiencies, 
eventually even leading to the loss of the cooling function.
However, when $\beta_\alpha \sqrt{\Gamma\delta} \gg 1$, $\epsilon+\sqrt{\Gamma\delta} \gg \mu_L$, and $\epsilon-\sqrt{\Gamma\delta} \ll \mu_R$, the two shifted excitation energies
will have left the transport window and will no longer participate in transport.
Then, tight coupling is again restored as only one transition energy $\Delta E = \epsilon$ remains inside the transport window.
We illustrate this in Fig.~\ref{FIG:transportwindow}.

Beyond a benchmark of the reaction coordinate treatment, this example provides a simple explanation of the observed behaviour and demonstrates that the mapping can be used as a 
tool to identify interesting parameter regimes, allowing for useful operational modes.
Strong coupling is not always detrimental to the efficiency of continuous heat engines.
However, regarding overall power output, it should be noted that we had to choose small $\delta$ (highly structured reservoirs) 
to maintain the validity of the reaction coordinate treatment.
This does of course bound the total power output of the device, which is not proportional to the coupling strength
in this regime.


\section{Outlook}

The reaction coordinate mapping is typically used to explore the strong-coupling regime, and as such it has found widespread application.
We stress here that it may also be extended to fermionic reservoirs.
By construction, the mapping itself is exact and thus barely suffers from additional constraints.
It can also be combined, for example, with formally exact methods such as nonequilibrium Green's functions~\cite{haug2008}, to simplify the
structure of the resulting equations.
However, when it is combined with some perturbative technique, its applicability will be limited to a certain degree.
For example, to apply a master equation to the supersystem, it is required that the residual (mapped) spectral density
allows for a Markovian treatment.
Thus, the master equation solution obtained via the reaction coordinate mapping will in general have a different range of validity 
than the master equation solution of the original system~\cite{iles_smith2014a,iles_smith2016a}.
We further note that with the mapping, one can transfer the time-dependence of parameters into the supersystem, thus enabling the treatment
of open-loop control schemes such as periodic driving~\cite{restrepo2018a}, or feedback control schemes such as Maxwell's demon~\cite{schaller2018a}, 
from the perspective of a driven system only.

The intuitive simplicity of the reaction coordinate technique makes it suitable to extend the range of validity of many different perturbative approaches. Indeed,  
beyond strong coupling, other problems can be treated with reaction coordinate mappings. For example, with the mapping one can study the effect of initial system-reservoir correlations by means of 
master equations.
Furthermore, as the treatment of the supersystem is Markovian, but the reduced dynamics of the original need not be, one can see the reaction coordinate mapping
as a Markovian embedding to study non-Markovian dynamics.

Finally, we mention that one can also use the mapping to engineer structured reservoirs.
Equipping a quantum system of interest with auxiliary degrees of freedom, which are coupled to structureless reservoirs, we can interpret the
auxiliary degrees of freedom as reaction coordinates and perform the inverse mapping.
Eventually, this yields a system coupled to structured reservoirs, with reaction coordinates that can for example be used as frequency filters to 
optimize performance of heat engines or other devices.

\section{Acknowledgments}

G.S. gratefully acknowledges discussions with J. Cerrillo, {N. Martensen}, S. Restrepo, and P. Strasberg and financial 
support by the DFG (GRK 1558, SFB 910, SCHA 1646/3-1, BR 1528/9-1). 

A.N. would like to thank D. Newman, {F. Mintert}, J. Iles-Smith, {N. Lambert}, Z. Blunden-Codd and V. Jouffrey for discussions. 
A.N. is supported by the Engineering and Physical Sciences Research Council, grant no. EP/N008154/1.


\bibliographystyle{apsrev4-1}

\begin{thebibliography}{57}%
\makeatletter
\providecommand \@ifxundefined [1]{%
 \@ifx{#1\undefined}
}%
\providecommand \@ifnum [1]{%
 \ifnum #1\expandafter \@firstoftwo
 \else \expandafter \@secondoftwo
 \fi
}%
\providecommand \@ifx [1]{%
 \ifx #1\expandafter \@firstoftwo
 \else \expandafter \@secondoftwo
 \fi
}%
\providecommand \natexlab [1]{#1}%
\providecommand \enquote  [1]{``#1''}%
\providecommand \bibnamefont  [1]{#1}%
\providecommand \bibfnamefont [1]{#1}%
\providecommand \citenamefont [1]{#1}%
\providecommand \href@noop [0]{\@secondoftwo}%
\providecommand \href [0]{\begingroup \@sanitize@url \@href}%
\providecommand \@href[1]{\@@startlink{#1}\@@href}%
\providecommand \@@href[1]{\endgroup#1\@@endlink}%
\providecommand \@sanitize@url [0]{\catcode `\\12\catcode `\$12\catcode
  `\&12\catcode `\#12\catcode `\^12\catcode `\_12\catcode `\%12\relax}%
\providecommand \@@startlink[1]{}%
\providecommand \@@endlink[0]{}%
\providecommand \url  [0]{\begingroup\@sanitize@url \@url }%
\providecommand \@url [1]{\endgroup\@href {#1}{\urlprefix }}%
\providecommand \urlprefix  [0]{URL }%
\providecommand \Eprint [0]{\href }%
\providecommand \doibase [0]{http://dx.doi.org/}%
\providecommand \selectlanguage [0]{\@gobble}%
\providecommand \bibinfo  [0]{\@secondoftwo}%
\providecommand \bibfield  [0]{\@secondoftwo}%
\providecommand \translation [1]{[#1]}%
\providecommand \BibitemOpen [0]{}%
\providecommand \bibitemStop [0]{}%
\providecommand \bibitemNoStop [0]{.\EOS\space}%
\providecommand \EOS [0]{\spacefactor3000\relax}%
\providecommand \BibitemShut  [1]{\csname bibitem#1\endcsname}%
\let\auto@bib@innerbib\@empty
\bibitem [{\citenamefont {Liu}\ \emph {et~al.}(2007)\citenamefont {Liu},
  \citenamefont {Zheng}, \citenamefont {Gong}, \citenamefont {Gao},\ and\
  \citenamefont {L\"u}}]{liu2007a}%
  \BibitemOpen
  \bibfield  {author} {\bibinfo {author} {\bibfnamefont {Y.}~\bibnamefont
  {Liu}}, \bibinfo {author} {\bibfnamefont {Y.}~\bibnamefont {Zheng}}, \bibinfo
  {author} {\bibfnamefont {W.}~\bibnamefont {Gong}}, \bibinfo {author}
  {\bibfnamefont {W.}~\bibnamefont {Gao}}, \ and\ \bibinfo {author}
  {\bibfnamefont {T.}~\bibnamefont {L\"u}},\ }\href {\doibase
  10.1016/j.physleta.2007.02.005} {\bibfield  {journal} {\bibinfo  {journal}
  {Physics Letters A}\ }\textbf {\bibinfo {volume} {365}},\ \bibinfo {pages}
  {495} (\bibinfo {year} {2007})}\BibitemShut {NoStop}%
\bibitem [{\citenamefont {Nesi}\ \emph {et~al.}(2007)\citenamefont {Nesi},
  \citenamefont {Paladino}, \citenamefont {Thorwart},\ and\ \citenamefont
  {Grifoni}}]{nesi2007b}%
  \BibitemOpen
  \bibfield  {author} {\bibinfo {author} {\bibfnamefont {F.}~\bibnamefont
  {Nesi}}, \bibinfo {author} {\bibfnamefont {E.}~\bibnamefont {Paladino}},
  \bibinfo {author} {\bibfnamefont {M.}~\bibnamefont {Thorwart}}, \ and\
  \bibinfo {author} {\bibfnamefont {M.}~\bibnamefont {Grifoni}},\ }\href@noop
  {} {\bibfield  {journal} {\bibinfo  {journal} {Europhysics Letters}\ }\textbf
  {\bibinfo {volume} {80}},\ \bibinfo {pages} {40005} (\bibinfo {year}
  {2007})}\BibitemShut {NoStop}%
\bibitem [{\citenamefont {H{\"o}rhammer}\ and\ \citenamefont
  {B{\"u}ttner}(2008)}]{Horhammer2008}%
  \BibitemOpen
  \bibfield  {author} {\bibinfo {author} {\bibfnamefont {C.}~\bibnamefont
  {H{\"o}rhammer}}\ and\ \bibinfo {author} {\bibfnamefont {H.}~\bibnamefont
  {B{\"u}ttner}},\ }\href {\doibase 10.1007/s10955-008-9640-x} {\bibfield
  {journal} {\bibinfo  {journal} {Journal of Statistical Physics}\ }\textbf
  {\bibinfo {volume} {133}},\ \bibinfo {pages} {1161} (\bibinfo {year}
  {2008})}\BibitemShut {NoStop}%
\bibitem [{\citenamefont {Campisi}\ \emph {et~al.}(2009)\citenamefont
  {Campisi}, \citenamefont {Talkner},\ and\ \citenamefont
  {H\"anggi}}]{campisi2009a}%
  \BibitemOpen
  \bibfield  {author} {\bibinfo {author} {\bibfnamefont {M.}~\bibnamefont
  {Campisi}}, \bibinfo {author} {\bibfnamefont {P.}~\bibnamefont {Talkner}}, \
  and\ \bibinfo {author} {\bibfnamefont {P.}~\bibnamefont {H\"anggi}},\ }\href
  {\doibase 10.1103/PhysRevLett.102.210401} {\bibfield  {journal} {\bibinfo
  {journal} {Phys. Rev. Lett.}\ }\textbf {\bibinfo {volume} {102}},\ \bibinfo
  {pages} {210401} (\bibinfo {year} {2009})}\BibitemShut {NoStop}%
\bibitem [{\citenamefont {Nicolin}\ and\ \citenamefont
  {Segal}(2011)}]{nicolin2011a}%
  \BibitemOpen
  \bibfield  {author} {\bibinfo {author} {\bibfnamefont {L.}~\bibnamefont
  {Nicolin}}\ and\ \bibinfo {author} {\bibfnamefont {D.}~\bibnamefont
  {Segal}},\ }\href {\doibase 10.1103/PhysRevB.84.161414} {\bibfield  {journal}
  {\bibinfo  {journal} {Physical Review B}\ }\textbf {\bibinfo {volume} {84}},\
  \bibinfo {pages} {161414} (\bibinfo {year} {2011})}\BibitemShut {NoStop}%
\bibitem [{\citenamefont {Deffner}\ and\ \citenamefont
  {Lutz}(2011)}]{deffner2011a}%
  \BibitemOpen
  \bibfield  {author} {\bibinfo {author} {\bibfnamefont {S.}~\bibnamefont
  {Deffner}}\ and\ \bibinfo {author} {\bibfnamefont {E.}~\bibnamefont {Lutz}},\
  }\href {\doibase 10.1103/PhysRevLett.107.140404} {\bibfield  {journal}
  {\bibinfo  {journal} {Phys. Rev. Lett.}\ }\textbf {\bibinfo {volume} {107}},\
  \bibinfo {pages} {140404} (\bibinfo {year} {2011})}\BibitemShut {NoStop}%
\bibitem [{\citenamefont {Hausinger}\ and\ \citenamefont
  {Grifoni}(2011)}]{hausinger2011a}%
  \BibitemOpen
  \bibfield  {author} {\bibinfo {author} {\bibfnamefont {J.}~\bibnamefont
  {Hausinger}}\ and\ \bibinfo {author} {\bibfnamefont {M.}~\bibnamefont
  {Grifoni}},\ }\href {\doibase 10.1103/PhysRevA.83.030301} {\bibfield
  {journal} {\bibinfo  {journal} {Physical Review A}\ }\textbf {\bibinfo
  {volume} {83}},\ \bibinfo {pages} {030301} (\bibinfo {year}
  {2011})}\BibitemShut {NoStop}%
\bibitem [{\citenamefont {Pucci}\ \emph {et~al.}(2013)\citenamefont {Pucci},
  \citenamefont {Esposito},\ and\ \citenamefont {Peliti}}]{pucci2013a}%
  \BibitemOpen
  \bibfield  {author} {\bibinfo {author} {\bibfnamefont {L.}~\bibnamefont
  {Pucci}}, \bibinfo {author} {\bibfnamefont {M.}~\bibnamefont {Esposito}}, \
  and\ \bibinfo {author} {\bibfnamefont {L.}~\bibnamefont {Peliti}},\ }\href
  {http://stacks.iop.org/1742-5468/2013/i=04/a=P04005} {\bibfield  {journal}
  {\bibinfo  {journal} {Journal of Statistical Mechanics: Theory and
  Experiment}\ }\textbf {\bibinfo {volume} {2013}},\ \bibinfo {pages} {P04005}
  (\bibinfo {year} {2013})}\BibitemShut {NoStop}%
\bibitem [{\citenamefont {Schaller}\ \emph {et~al.}(2013)\citenamefont
  {Schaller}, \citenamefont {Krause}, \citenamefont {Brandes},\ and\
  \citenamefont {Esposito}}]{schaller2013a}%
  \BibitemOpen
  \bibfield  {author} {\bibinfo {author} {\bibfnamefont {G.}~\bibnamefont
  {Schaller}}, \bibinfo {author} {\bibfnamefont {T.}~\bibnamefont {Krause}},
  \bibinfo {author} {\bibfnamefont {T.}~\bibnamefont {Brandes}}, \ and\
  \bibinfo {author} {\bibfnamefont {M.}~\bibnamefont {Esposito}},\ }\href
  {\doibase 10.1088/1367-2630/15/3/033032} {\bibfield  {journal} {\bibinfo
  {journal} {New Journal of Physics}\ }\textbf {\bibinfo {volume} {15}},\
  \bibinfo {pages} {033032} (\bibinfo {year} {2013})}\BibitemShut {NoStop}%
\bibitem [{\citenamefont {Ankerhold}\ and\ \citenamefont
  {Pekola}(2014)}]{ankerhold2014a}%
  \BibitemOpen
  \bibfield  {author} {\bibinfo {author} {\bibfnamefont {J.}~\bibnamefont
  {Ankerhold}}\ and\ \bibinfo {author} {\bibfnamefont {J.~P.}\ \bibnamefont
  {Pekola}},\ }\href {\doibase 10.1103/PhysRevB.90.075421} {\bibfield
  {journal} {\bibinfo  {journal} {Phys. Rev. B}\ }\textbf {\bibinfo {volume}
  {90}},\ \bibinfo {pages} {075421} (\bibinfo {year} {2014})}\BibitemShut
  {NoStop}%
\bibitem [{\citenamefont {Iles-Smith}\ \emph {et~al.}(2014)\citenamefont
  {Iles-Smith}, \citenamefont {Lambert},\ and\ \citenamefont
  {Nazir}}]{iles_smith2014a}%
  \BibitemOpen
  \bibfield  {author} {\bibinfo {author} {\bibfnamefont {J.}~\bibnamefont
  {Iles-Smith}}, \bibinfo {author} {\bibfnamefont {N.}~\bibnamefont {Lambert}},
  \ and\ \bibinfo {author} {\bibfnamefont {A.}~\bibnamefont {Nazir}},\ }\href
  {\doibase 10.1103/PhysRevA.90.032114} {\bibfield  {journal} {\bibinfo
  {journal} {Physical Review A}\ }\textbf {\bibinfo {volume} {90}},\ \bibinfo
  {pages} {032114} (\bibinfo {year} {2014})}\BibitemShut {NoStop}%
\bibitem [{\citenamefont {Gallego}\ \emph {et~al.}(2014)\citenamefont
  {Gallego}, \citenamefont {Riera},\ and\ \citenamefont
  {Eisert}}]{gallego2014a}%
  \BibitemOpen
  \bibfield  {author} {\bibinfo {author} {\bibfnamefont {R.}~\bibnamefont
  {Gallego}}, \bibinfo {author} {\bibfnamefont {A.}~\bibnamefont {Riera}}, \
  and\ \bibinfo {author} {\bibfnamefont {J.}~\bibnamefont {Eisert}},\ }\href
  {http://stacks.iop.org/1367-2630/16/i=12/a=125009} {\bibfield  {journal}
  {\bibinfo  {journal} {New Journal of Physics}\ }\textbf {\bibinfo {volume}
  {16}},\ \bibinfo {pages} {125009} (\bibinfo {year} {2014})}\BibitemShut
  {NoStop}%
\bibitem [{\citenamefont {Wang}\ \emph {et~al.}(2015)\citenamefont {Wang},
  \citenamefont {Ren},\ and\ \citenamefont {Cao}}]{wang2015a}%
  \BibitemOpen
  \bibfield  {author} {\bibinfo {author} {\bibfnamefont {C.}~\bibnamefont
  {Wang}}, \bibinfo {author} {\bibfnamefont {J.}~\bibnamefont {Ren}}, \ and\
  \bibinfo {author} {\bibfnamefont {J.}~\bibnamefont {Cao}},\ }\href {\doibase
  10.1038/srep11787} {\bibfield  {journal} {\bibinfo  {journal} {Scientific
  Reports}\ }\textbf {\bibinfo {volume} {5}},\ \bibinfo {pages} {11787}
  (\bibinfo {year} {2015})}\BibitemShut {NoStop}%
\bibitem [{\citenamefont {Esposito}\ \emph
  {et~al.}(2015{\natexlab{a}})\citenamefont {Esposito}, \citenamefont {Ochoa},\
  and\ \citenamefont {Galperin}}]{esposito2015a}%
  \BibitemOpen
  \bibfield  {author} {\bibinfo {author} {\bibfnamefont {M.}~\bibnamefont
  {Esposito}}, \bibinfo {author} {\bibfnamefont {M.~A.}\ \bibnamefont {Ochoa}},
  \ and\ \bibinfo {author} {\bibfnamefont {M.}~\bibnamefont {Galperin}},\
  }\href {\doibase 10.1103/PhysRevLett.114.080602} {\bibfield  {journal}
  {\bibinfo  {journal} {Phys. Rev. Lett.}\ }\textbf {\bibinfo {volume} {114}},\
  \bibinfo {pages} {080602} (\bibinfo {year} {2015}{\natexlab{a}})}\BibitemShut
  {NoStop}%
\bibitem [{\citenamefont {Esposito}\ \emph
  {et~al.}(2015{\natexlab{b}})\citenamefont {Esposito}, \citenamefont {Ochoa},\
  and\ \citenamefont {Galperin}}]{esposito2015b}%
  \BibitemOpen
  \bibfield  {author} {\bibinfo {author} {\bibfnamefont {M.}~\bibnamefont
  {Esposito}}, \bibinfo {author} {\bibfnamefont {M.~A.}\ \bibnamefont {Ochoa}},
  \ and\ \bibinfo {author} {\bibfnamefont {M.}~\bibnamefont {Galperin}},\
  }\href {\doibase 10.1103/PhysRevB.92.235440} {\bibfield  {journal} {\bibinfo
  {journal} {Physical Review B}\ }\textbf {\bibinfo {volume} {92}},\ \bibinfo
  {pages} {235440} (\bibinfo {year} {2015}{\natexlab{b}})}\BibitemShut
  {NoStop}%
\bibitem [{\citenamefont {Gelbwaser-Klimovsky}\ and\ \citenamefont
  {Aspuru-Guzik}(2015)}]{gelbwaser_klimovsky2015a}%
  \BibitemOpen
  \bibfield  {author} {\bibinfo {author} {\bibfnamefont {D.}~\bibnamefont
  {Gelbwaser-Klimovsky}}\ and\ \bibinfo {author} {\bibfnamefont
  {A.}~\bibnamefont {Aspuru-Guzik}},\ }\href {\doibase
  10.1021/acs.jpclett.5b01404} {\bibfield  {journal} {\bibinfo  {journal} {The
  Journal of Physical Chemistry Letters}\ }\textbf {\bibinfo {volume} {6}},\
  \bibinfo {pages} {3477} (\bibinfo {year} {2015})}\BibitemShut {NoStop}%
\bibitem [{\citenamefont {Carrega}\ \emph {et~al.}(2015)\citenamefont
  {Carrega}, \citenamefont {Solinas}, \citenamefont {Braggio}, \citenamefont
  {Sassetti},\ and\ \citenamefont {Weiss}}]{carrega2015a}%
  \BibitemOpen
  \bibfield  {author} {\bibinfo {author} {\bibfnamefont {M.}~\bibnamefont
  {Carrega}}, \bibinfo {author} {\bibfnamefont {P.}~\bibnamefont {Solinas}},
  \bibinfo {author} {\bibfnamefont {A.}~\bibnamefont {Braggio}}, \bibinfo
  {author} {\bibfnamefont {M.}~\bibnamefont {Sassetti}}, \ and\ \bibinfo
  {author} {\bibfnamefont {U.}~\bibnamefont {Weiss}},\ }\href
  {http://stacks.iop.org/1367-2630/17/i=4/a=045030} {\bibfield  {journal}
  {\bibinfo  {journal} {New Journal of Physics}\ }\textbf {\bibinfo {volume}
  {17}},\ \bibinfo {pages} {045030} (\bibinfo {year} {2015})}\BibitemShut
  {NoStop}%
\bibitem [{\citenamefont {Strasberg}\ \emph {et~al.}(2016)\citenamefont
  {Strasberg}, \citenamefont {Schaller}, \citenamefont {Lambert},\ and\
  \citenamefont {Brandes}}]{strasberg2016a}%
  \BibitemOpen
  \bibfield  {author} {\bibinfo {author} {\bibfnamefont {P.}~\bibnamefont
  {Strasberg}}, \bibinfo {author} {\bibfnamefont {G.}~\bibnamefont {Schaller}},
  \bibinfo {author} {\bibfnamefont {N.}~\bibnamefont {Lambert}}, \ and\
  \bibinfo {author} {\bibfnamefont {T.}~\bibnamefont {Brandes}},\ }\href
  {\doibase 10.1088/1367-2630/18/7/073007} {\bibfield  {journal} {\bibinfo
  {journal} {New Journal of Physics}\ }\textbf {\bibinfo {volume} {18}},\
  \bibinfo {pages} {073007} (\bibinfo {year} {2016})}\BibitemShut {NoStop}%
\bibitem [{\citenamefont {Katz}\ and\ \citenamefont
  {Kosloff}(2016)}]{kac2016a}%
  \BibitemOpen
  \bibfield  {author} {\bibinfo {author} {\bibfnamefont {G.}~\bibnamefont
  {Katz}}\ and\ \bibinfo {author} {\bibfnamefont {R.}~\bibnamefont {Kosloff}},\
  }\href {\doibase 10.3390/e18050186} {\bibfield  {journal} {\bibinfo
  {journal} {Entropy}\ }\textbf {\bibinfo {volume} {18}},\ \bibinfo {pages}
  {186} (\bibinfo {year} {2016})}\BibitemShut {NoStop}%
\bibitem [{\citenamefont {Cerrillo}\ \emph {et~al.}(2016)\citenamefont
  {Cerrillo}, \citenamefont {Buser},\ and\ \citenamefont
  {Brandes}}]{cerrillo2016a}%
  \BibitemOpen
  \bibfield  {author} {\bibinfo {author} {\bibfnamefont {J.}~\bibnamefont
  {Cerrillo}}, \bibinfo {author} {\bibfnamefont {M.}~\bibnamefont {Buser}}, \
  and\ \bibinfo {author} {\bibfnamefont {T.}~\bibnamefont {Brandes}},\ }\href
  {\doibase 10.1103/PhysRevB.94.214308} {\bibfield  {journal} {\bibinfo
  {journal} {Physical Review B}\ }\textbf {\bibinfo {volume} {94}},\ \bibinfo
  {pages} {214308} (\bibinfo {year} {2016})}\BibitemShut {NoStop}%
\bibitem [{\citenamefont {Seifert}(2016)}]{seifert2016a}%
  \BibitemOpen
  \bibfield  {author} {\bibinfo {author} {\bibfnamefont {U.}~\bibnamefont
  {Seifert}},\ }\href {\doibase 10.1103/PhysRevLett.116.020601} {\bibfield
  {journal} {\bibinfo  {journal} {Phys. Rev. Lett.}\ }\textbf {\bibinfo
  {volume} {116}},\ \bibinfo {pages} {020601} (\bibinfo {year}
  {2016})}\BibitemShut {NoStop}%
\bibitem [{\citenamefont {Newman}\ \emph {et~al.}(2017)\citenamefont {Newman},
  \citenamefont {Mintert},\ and\ \citenamefont {Nazir}}]{newman2017a}%
  \BibitemOpen
  \bibfield  {author} {\bibinfo {author} {\bibfnamefont {D.}~\bibnamefont
  {Newman}}, \bibinfo {author} {\bibfnamefont {F.}~\bibnamefont {Mintert}}, \
  and\ \bibinfo {author} {\bibfnamefont {A.}~\bibnamefont {Nazir}},\ }\href
  {\doibase 10.1103/PhysRevE.95.032139} {\bibfield  {journal} {\bibinfo
  {journal} {Physical Review E}\ }\textbf {\bibinfo {volume} {95}},\ \bibinfo
  {pages} {032139} (\bibinfo {year} {2017})}\BibitemShut {NoStop}%
\bibitem [{\citenamefont {Strasberg}\ and\ \citenamefont
  {Esposito}(2017)}]{strasberg2017b}%
  \BibitemOpen
  \bibfield  {author} {\bibinfo {author} {\bibfnamefont {P.}~\bibnamefont
  {Strasberg}}\ and\ \bibinfo {author} {\bibfnamefont {M.}~\bibnamefont
  {Esposito}},\ }\href {\doibase 10.1103/PhysRevE.95.062101} {\bibfield
  {journal} {\bibinfo  {journal} {Physical Review E}\ }\textbf {\bibinfo
  {volume} {95}},\ \bibinfo {pages} {062101} (\bibinfo {year}
  {2017})}\BibitemShut {NoStop}%
\bibitem [{\citenamefont {Miller}\ and\ \citenamefont
  {Anders}(2017)}]{miller2017a}%
  \BibitemOpen
  \bibfield  {author} {\bibinfo {author} {\bibfnamefont {H.~J.~D.}\
  \bibnamefont {Miller}}\ and\ \bibinfo {author} {\bibfnamefont
  {J.}~\bibnamefont {Anders}},\ }\href {\doibase 10.1103/PhysRevE.95.062123}
  {\bibfield  {journal} {\bibinfo  {journal} {Phys. Rev. E}\ }\textbf {\bibinfo
  {volume} {95}},\ \bibinfo {pages} {062123} (\bibinfo {year}
  {2017})}\BibitemShut {NoStop}%
\bibitem [{\citenamefont {Mu}\ \emph {et~al.}(2017)\citenamefont {Mu},
  \citenamefont {Agarwalla}, \citenamefont {Schaller},\ and\ \citenamefont
  {Segal}}]{mu2017a}%
  \BibitemOpen
  \bibfield  {author} {\bibinfo {author} {\bibfnamefont {A.}~\bibnamefont
  {Mu}}, \bibinfo {author} {\bibfnamefont {B.~K.}\ \bibnamefont {Agarwalla}},
  \bibinfo {author} {\bibfnamefont {G.}~\bibnamefont {Schaller}}, \ and\
  \bibinfo {author} {\bibfnamefont {D.}~\bibnamefont {Segal}},\ }\href
  {\doibase 10.1088/1367-2630/aa9b75} {\bibfield  {journal} {\bibinfo
  {journal} {New Journal of Physics}\ }\textbf {\bibinfo {volume} {19}},\
  \bibinfo {pages} {123034} (\bibinfo {year} {2017})}\BibitemShut {NoStop}%
\bibitem [{\citenamefont {Jarzynski}(2017)}]{jarzynski2017a}%
  \BibitemOpen
  \bibfield  {author} {\bibinfo {author} {\bibfnamefont {C.}~\bibnamefont
  {Jarzynski}},\ }\href {\doibase 10.1103/PhysRevX.7.011008} {\bibfield
  {journal} {\bibinfo  {journal} {Phys. Rev. X}\ }\textbf {\bibinfo {volume}
  {7}},\ \bibinfo {pages} {011008} (\bibinfo {year} {2017})}\BibitemShut
  {NoStop}%
\bibitem [{\citenamefont {Freitas}\ and\ \citenamefont
  {Paz}(2017)}]{freitas2017a}%
  \BibitemOpen
  \bibfield  {author} {\bibinfo {author} {\bibfnamefont {N.}~\bibnamefont
  {Freitas}}\ and\ \bibinfo {author} {\bibfnamefont {J.~P.}\ \bibnamefont
  {Paz}},\ }\href {\doibase 10.1103/PhysRevE.95.012146} {\bibfield  {journal}
  {\bibinfo  {journal} {Phys. Rev. E}\ }\textbf {\bibinfo {volume} {95}},\
  \bibinfo {pages} {012146} (\bibinfo {year} {2017})}\BibitemShut {NoStop}%
\bibitem [{\citenamefont {Perarnau-Llobet}\ \emph {et~al.}(2018)\citenamefont
  {Perarnau-Llobet}, \citenamefont {Wilming}, \citenamefont {Riera},
  \citenamefont {Gallego},\ and\ \citenamefont
  {Eisert}}]{perarnau_llobet2018a}%
  \BibitemOpen
  \bibfield  {author} {\bibinfo {author} {\bibfnamefont {M.}~\bibnamefont
  {Perarnau-Llobet}}, \bibinfo {author} {\bibfnamefont {H.}~\bibnamefont
  {Wilming}}, \bibinfo {author} {\bibfnamefont {A.}~\bibnamefont {Riera}},
  \bibinfo {author} {\bibfnamefont {R.}~\bibnamefont {Gallego}}, \ and\
  \bibinfo {author} {\bibfnamefont {J.}~\bibnamefont {Eisert}},\ }\href
  {\doibase 10.1103/PhysRevLett.120.120602} {\bibfield  {journal} {\bibinfo
  {journal} {Phys. Rev. Lett.}\ }\textbf {\bibinfo {volume} {120}},\ \bibinfo
  {pages} {120602} (\bibinfo {year} {2018})}\BibitemShut {NoStop}%
\bibitem [{\citenamefont {Burkey}\ and\ \citenamefont
  {Cantrell}(1984)}]{burkey1984a}%
  \BibitemOpen
  \bibfield  {author} {\bibinfo {author} {\bibfnamefont {R.~S.}\ \bibnamefont
  {Burkey}}\ and\ \bibinfo {author} {\bibfnamefont {C.~D.}\ \bibnamefont
  {Cantrell}},\ }\href {\doibase 10.1364/JOSAB.1.000169} {\bibfield  {journal}
  {\bibinfo  {journal} {Journal of the Optical Society of America B}\ }\textbf
  {\bibinfo {volume} {1}},\ \bibinfo {pages} {169} (\bibinfo {year}
  {1984})}\BibitemShut {NoStop}%
\bibitem [{\citenamefont {Garg}\ \emph {et~al.}(1985)\citenamefont {Garg},
  \citenamefont {Onuchic},\ and\ \citenamefont {Ambegaokar}}]{garg1985a}%
  \BibitemOpen
  \bibfield  {author} {\bibinfo {author} {\bibfnamefont {A.}~\bibnamefont
  {Garg}}, \bibinfo {author} {\bibfnamefont {J.~N.}\ \bibnamefont {Onuchic}}, \
  and\ \bibinfo {author} {\bibfnamefont {V.}~\bibnamefont {Ambegaokar}},\
  }\href {\doibase 10.1063/1.449017} {\bibfield  {journal} {\bibinfo  {journal}
  {The Journal of Chemical Physics}\ }\textbf {\bibinfo {volume} {83}},\
  \bibinfo {pages} {4491} (\bibinfo {year} {1985})}\BibitemShut {NoStop}%
\bibitem [{\citenamefont {Martinazzo}\ \emph {et~al.}(2011)\citenamefont
  {Martinazzo}, \citenamefont {Vacchini}, \citenamefont {Hughes},\ and\
  \citenamefont {Burghardt}}]{martinazzo2011a}%
  \BibitemOpen
  \bibfield  {author} {\bibinfo {author} {\bibfnamefont {R.}~\bibnamefont
  {Martinazzo}}, \bibinfo {author} {\bibfnamefont {B.}~\bibnamefont
  {Vacchini}}, \bibinfo {author} {\bibfnamefont {K.~H.}\ \bibnamefont
  {Hughes}}, \ and\ \bibinfo {author} {\bibfnamefont {I.}~\bibnamefont
  {Burghardt}},\ }\href {\doibase 10.1063/1.3532408} {\bibfield  {journal}
  {\bibinfo  {journal} {The Journal of Chemical Physics}\ }\textbf {\bibinfo
  {volume} {134}},\ \bibinfo {pages} {011101} (\bibinfo {year}
  {2011})}\BibitemShut {NoStop}%
\bibitem [{\citenamefont {Woods}\ \emph {et~al.}(2014)\citenamefont {Woods},
  \citenamefont {Groux}, \citenamefont {Chin}, \citenamefont {Huelga},\ and\
  \citenamefont {Plenio}}]{woods2014a}%
  \BibitemOpen
  \bibfield  {author} {\bibinfo {author} {\bibfnamefont {M.~P.}\ \bibnamefont
  {Woods}}, \bibinfo {author} {\bibfnamefont {R.}~\bibnamefont {Groux}},
  \bibinfo {author} {\bibfnamefont {A.~W.}\ \bibnamefont {Chin}}, \bibinfo
  {author} {\bibfnamefont {S.~F.}\ \bibnamefont {Huelga}}, \ and\ \bibinfo
  {author} {\bibfnamefont {M.~B.}\ \bibnamefont {Plenio}},\ }\href {\doibase
  10.1063/1.4866769} {\bibfield  {journal} {\bibinfo  {journal} {Journal of
  Mathematical Physics}\ }\textbf {\bibinfo {volume} {55}},\ \bibinfo {pages}
  {032101} (\bibinfo {year} {2014})}\BibitemShut {NoStop}%
\bibitem [{\citenamefont {Schaller}\ \emph {et~al.}(2018)\citenamefont
  {Schaller}, \citenamefont {Cerrillo}, \citenamefont {Engelhardt},\ and\
  \citenamefont {Strasberg}}]{schaller2018a}%
  \BibitemOpen
  \bibfield  {author} {\bibinfo {author} {\bibfnamefont {G.}~\bibnamefont
  {Schaller}}, \bibinfo {author} {\bibfnamefont {J.}~\bibnamefont {Cerrillo}},
  \bibinfo {author} {\bibfnamefont {G.}~\bibnamefont {Engelhardt}}, \ and\
  \bibinfo {author} {\bibfnamefont {P.}~\bibnamefont {Strasberg}},\ }\href
  {\doibase 10.1103/PhysRevB.97.195104} {\bibfield  {journal} {\bibinfo
  {journal} {Phys. Rev. B}\ }\textbf {\bibinfo {volume} {97}},\ \bibinfo
  {pages} {195104} (\bibinfo {year} {2018})}\BibitemShut {NoStop}%
\bibitem [{\citenamefont {Strasberg}\ \emph {et~al.}(2018)\citenamefont
  {Strasberg}, \citenamefont {Schaller}, \citenamefont {Schmidt},\ and\
  \citenamefont {Esposito}}]{strasberg2018a}%
  \BibitemOpen
  \bibfield  {author} {\bibinfo {author} {\bibfnamefont {P.}~\bibnamefont
  {Strasberg}}, \bibinfo {author} {\bibfnamefont {G.}~\bibnamefont {Schaller}},
  \bibinfo {author} {\bibfnamefont {T.~L.}\ \bibnamefont {Schmidt}}, \ and\
  \bibinfo {author} {\bibfnamefont {M.}~\bibnamefont {Esposito}},\ }\href
  {\doibase 10.1103/PhysRevB.97.205405} {\bibfield  {journal} {\bibinfo
  {journal} {Phys. Rev. B}\ }\textbf {\bibinfo {volume} {97}},\ \bibinfo
  {pages} {205405} (\bibinfo {year} {2018})}\BibitemShut {NoStop}%
\bibitem [{\citenamefont {Restrepo}\ \emph {et~al.}(2018)\citenamefont
  {Restrepo}, \citenamefont {Cerrillo}, \citenamefont {Strasberg},\ and\
  \citenamefont {Schaller}}]{restrepo2018a}%
  \BibitemOpen
  \bibfield  {author} {\bibinfo {author} {\bibfnamefont {S.}~\bibnamefont
  {Restrepo}}, \bibinfo {author} {\bibfnamefont {J.}~\bibnamefont {Cerrillo}},
  \bibinfo {author} {\bibfnamefont {P.}~\bibnamefont {Strasberg}}, \ and\
  \bibinfo {author} {\bibfnamefont {G.}~\bibnamefont {Schaller}},\ }\href
  {\doibase 10.1088/1367-2630/aac583} {\bibfield  {journal} {\bibinfo
  {journal} {New Journal of Physics}\ }\textbf {\bibinfo {volume}
  {doi:10.1088/1367-2630/aac583}} (\bibinfo {year} {2018}),\
  10.1088/1367-2630/aac583}\BibitemShut {NoStop}%
\bibitem [{\citenamefont {Iles-Smith}\ \emph {et~al.}(2016)\citenamefont
  {Iles-Smith}, \citenamefont {Dijkstra}, \citenamefont {Lambert},\ and\
  \citenamefont {Nazir}}]{iles_smith2016a}%
  \BibitemOpen
  \bibfield  {author} {\bibinfo {author} {\bibfnamefont {J.}~\bibnamefont
  {Iles-Smith}}, \bibinfo {author} {\bibfnamefont {A.~G.}\ \bibnamefont
  {Dijkstra}}, \bibinfo {author} {\bibfnamefont {N.}~\bibnamefont {Lambert}}, \
  and\ \bibinfo {author} {\bibfnamefont {A.}~\bibnamefont {Nazir}},\ }\href
  {\doibase 10.1063/1.4940218} {\bibfield  {journal} {\bibinfo  {journal} {The
  Journal of Chemical Physics}\ }\textbf {\bibinfo {volume} {144}},\ \bibinfo
  {pages} {044110} (\bibinfo {year} {2016})}\BibitemShut {NoStop}%
\bibitem [{\citenamefont {Huh}\ \emph {et~al.}(2014)\citenamefont {Huh},
  \citenamefont {Mostame}, \citenamefont {Fujita}, \citenamefont {Yung},\ and\
  \citenamefont {Aspuru-Guzik}}]{huh2014a}%
  \BibitemOpen
  \bibfield  {author} {\bibinfo {author} {\bibfnamefont {J.}~\bibnamefont
  {Huh}}, \bibinfo {author} {\bibfnamefont {S.}~\bibnamefont {Mostame}},
  \bibinfo {author} {\bibfnamefont {T.}~\bibnamefont {Fujita}}, \bibinfo
  {author} {\bibfnamefont {M.-H.}\ \bibnamefont {Yung}}, \ and\ \bibinfo
  {author} {\bibfnamefont {A.}~\bibnamefont {Aspuru-Guzik}},\ }\href {\doibase
  10.1088/1367-2630/16/12/123008} {\bibfield  {journal} {\bibinfo  {journal}
  {New Journal of Physics}\ }\textbf {\bibinfo {volume} {16}},\ \bibinfo
  {pages} {123008} (\bibinfo {year} {2014})}\BibitemShut {NoStop}%
\bibitem [{\citenamefont {Woods}\ \emph {et~al.}(2015)\citenamefont {Woods},
  \citenamefont {Cramer},\ and\ \citenamefont {Plenio}}]{woods2015a}%
  \BibitemOpen
  \bibfield  {author} {\bibinfo {author} {\bibfnamefont {M.~P.}\ \bibnamefont
  {Woods}}, \bibinfo {author} {\bibfnamefont {M.}~\bibnamefont {Cramer}}, \
  and\ \bibinfo {author} {\bibfnamefont {M.~B.}\ \bibnamefont {Plenio}},\
  }\href {\doibase 10.1103/PhysRevLett.115.130401} {\bibfield  {journal}
  {\bibinfo  {journal} {Physical Review Letters}\ }\textbf {\bibinfo {volume}
  {115}},\ \bibinfo {pages} {130401} (\bibinfo {year} {2015})}\BibitemShut
  {NoStop}%
\bibitem [{\citenamefont {Woods}\ and\ \citenamefont
  {Plenio}(2016)}]{woods2016a}%
  \BibitemOpen
  \bibfield  {author} {\bibinfo {author} {\bibfnamefont {M.~P.}\ \bibnamefont
  {Woods}}\ and\ \bibinfo {author} {\bibfnamefont {M.~B.}\ \bibnamefont
  {Plenio}},\ }\href {\doibase 10.1063/1.4940436} {\bibfield  {journal}
  {\bibinfo  {journal} {Journal of Mathematical Physics}\ }\textbf {\bibinfo
  {volume} {57}},\ \bibinfo {pages} {022105} (\bibinfo {year}
  {2016})}\BibitemShut {NoStop}%
\bibitem [{\citenamefont {Gogolin}\ and\ \citenamefont
  {Eisert}(2016)}]{gogolin2016a}%
  \BibitemOpen
  \bibfield  {author} {\bibinfo {author} {\bibfnamefont {C.}~\bibnamefont
  {Gogolin}}\ and\ \bibinfo {author} {\bibfnamefont {J.}~\bibnamefont
  {Eisert}},\ }\href {\doibase 10.1088/0034-4885/79/5/056001} {\bibfield
  {journal} {\bibinfo  {journal} {Reports on Progress in Physics}\ }\textbf
  {\bibinfo {volume} {79}},\ \bibinfo {pages} {056001} (\bibinfo {year}
  {2016})}\BibitemShut {NoStop}%
\bibitem [{\citenamefont {D\"umcke}\ and\ \citenamefont
  {Spohn}(1979)}]{duemcke1979a}%
  \BibitemOpen
  \bibfield  {author} {\bibinfo {author} {\bibfnamefont {R.}~\bibnamefont
  {D\"umcke}}\ and\ \bibinfo {author} {\bibfnamefont {H.}~\bibnamefont
  {Spohn}},\ }\href@noop {} {\bibfield  {journal} {\bibinfo  {journal}
  {Zeitschrift f\"ur Physik B}\ }\textbf {\bibinfo {volume} {34}},\ \bibinfo
  {pages} {419} (\bibinfo {year} {1979})}\BibitemShut {NoStop}%
\bibitem [{\citenamefont {Breuer}\ and\ \citenamefont
  {Petruccione}(2002)}]{breuer2002}%
  \BibitemOpen
  \bibfield  {author} {\bibinfo {author} {\bibfnamefont {H.-P.}\ \bibnamefont
  {Breuer}}\ and\ \bibinfo {author} {\bibfnamefont {F.}~\bibnamefont
  {Petruccione}},\ }\href@noop {} {\emph {\bibinfo {title} {The Theory of Open
  Quantum Systems}}}\ (\bibinfo  {publisher} {Oxford University Press},\
  \bibinfo {address} {Oxford},\ \bibinfo {year} {2002})\BibitemShut {NoStop}%
\bibitem [{\citenamefont {Le~Hur}(2012)}]{lehur2012a}%
  \BibitemOpen
  \bibfield  {author} {\bibinfo {author} {\bibfnamefont {K.}~\bibnamefont
  {Le~Hur}},\ }\href {\doibase 10.1103/PhysRevB.85.140506} {\bibfield
  {journal} {\bibinfo  {journal} {Phys. Rev. B}\ }\textbf {\bibinfo {volume}
  {85}},\ \bibinfo {pages} {140506} (\bibinfo {year} {2012})}\BibitemShut
  {NoStop}%
\bibitem [{\citenamefont {Goldstein}\ \emph {et~al.}(2013)\citenamefont
  {Goldstein}, \citenamefont {Devoret}, \citenamefont {Houzet},\ and\
  \citenamefont {Glazman}}]{goldstein2013a}%
  \BibitemOpen
  \bibfield  {author} {\bibinfo {author} {\bibfnamefont {M.}~\bibnamefont
  {Goldstein}}, \bibinfo {author} {\bibfnamefont {M.~H.}\ \bibnamefont
  {Devoret}}, \bibinfo {author} {\bibfnamefont {M.}~\bibnamefont {Houzet}}, \
  and\ \bibinfo {author} {\bibfnamefont {L.~I.}\ \bibnamefont {Glazman}},\
  }\href {\doibase 10.1103/PhysRevLett.110.017002} {\bibfield  {journal}
  {\bibinfo  {journal} {Phys. Rev. Lett.}\ }\textbf {\bibinfo {volume} {110}},\
  \bibinfo {pages} {017002} (\bibinfo {year} {2013})}\BibitemShut {NoStop}%
\bibitem [{\citenamefont {Peropadre}\ \emph {et~al.}(2013)\citenamefont
  {Peropadre}, \citenamefont {Zueco}, \citenamefont {Porras},\ and\
  \citenamefont {Garc\'{\i}a-Ripoll}}]{peropadre2013a}%
  \BibitemOpen
  \bibfield  {author} {\bibinfo {author} {\bibfnamefont {B.}~\bibnamefont
  {Peropadre}}, \bibinfo {author} {\bibfnamefont {D.}~\bibnamefont {Zueco}},
  \bibinfo {author} {\bibfnamefont {D.}~\bibnamefont {Porras}}, \ and\ \bibinfo
  {author} {\bibfnamefont {J.~J.}\ \bibnamefont {Garc\'{\i}a-Ripoll}},\ }\href
  {\doibase 10.1103/PhysRevLett.111.243602} {\bibfield  {journal} {\bibinfo
  {journal} {Phys. Rev. Lett.}\ }\textbf {\bibinfo {volume} {111}},\ \bibinfo
  {pages} {243602} (\bibinfo {year} {2013})}\BibitemShut {NoStop}%
\bibitem [{\citenamefont {Nazir}\ and\ \citenamefont
  {McCutcheon}(2016)}]{nazir2016a}%
  \BibitemOpen
  \bibfield  {author} {\bibinfo {author} {\bibfnamefont {A.}~\bibnamefont
  {Nazir}}\ and\ \bibinfo {author} {\bibfnamefont {D.~P.~S.}\ \bibnamefont
  {McCutcheon}},\ }\href {http://stacks.iop.org/0953-8984/28/i=10/a=103002}
  {\bibfield  {journal} {\bibinfo  {journal} {Journal of Physics: Condensed
  Matter}\ }\textbf {\bibinfo {volume} {28}},\ \bibinfo {pages} {103002}
  (\bibinfo {year} {2016})}\BibitemShut {NoStop}%
\bibitem [{\citenamefont {Kosloff}\ and\ \citenamefont
  {Levy}(2014)}]{kosloff2014a}%
  \BibitemOpen
  \bibfield  {author} {\bibinfo {author} {\bibfnamefont {R.}~\bibnamefont
  {Kosloff}}\ and\ \bibinfo {author} {\bibfnamefont {A.}~\bibnamefont {Levy}},\
  }\href {\doibase 10.1146/annurev-physchem-040513-103724} {\bibfield
  {journal} {\bibinfo  {journal} {Annual Review of Physical Chemistry}\
  }\textbf {\bibinfo {volume} {65}},\ \bibinfo {pages} {365} (\bibinfo {year}
  {2014})}\BibitemShut {NoStop}%
\bibitem [{\citenamefont {Esposito}\ \emph {et~al.}(2009)\citenamefont
  {Esposito}, \citenamefont {Lindenberg},\ and\ \citenamefont {den
  Broeck}}]{esposito2009b}%
  \BibitemOpen
  \bibfield  {author} {\bibinfo {author} {\bibfnamefont {M.}~\bibnamefont
  {Esposito}}, \bibinfo {author} {\bibfnamefont {K.}~\bibnamefont
  {Lindenberg}}, \ and\ \bibinfo {author} {\bibfnamefont {C.~V.}\ \bibnamefont
  {den Broeck}},\ }\href {\doibase 10.1209/0295-5075/85/60010} {\bibfield
  {journal} {\bibinfo  {journal} {Europhysics Letters}\ }\textbf {\bibinfo
  {volume} {85}},\ \bibinfo {pages} {60010} (\bibinfo {year}
  {2009})}\BibitemShut {NoStop}%
\bibitem [{\citenamefont {Haug}\ and\ \citenamefont {Jauho}(2008)}]{haug2008}%
  \BibitemOpen
  \bibfield  {author} {\bibinfo {author} {\bibfnamefont {H.}~\bibnamefont
  {Haug}}\ and\ \bibinfo {author} {\bibfnamefont {A.-P.}\ \bibnamefont
  {Jauho}},\ }\href@noop {} {\emph {\bibinfo {title} {Quantum Kinetics in
  Transport and Optics of Semiconductors}}}\ (\bibinfo  {publisher}
  {Springer},\ \bibinfo {year} {2008})\BibitemShut {NoStop}%
\bibitem [{\citenamefont {Topp}\ \emph {et~al.}(2015)\citenamefont {Topp},
  \citenamefont {Brandes},\ and\ \citenamefont {Schaller}}]{topp2015a}%
  \BibitemOpen
  \bibfield  {author} {\bibinfo {author} {\bibfnamefont {G.~E.}\ \bibnamefont
  {Topp}}, \bibinfo {author} {\bibfnamefont {T.}~\bibnamefont {Brandes}}, \
  and\ \bibinfo {author} {\bibfnamefont {G.}~\bibnamefont {Schaller}},\ }\href
  {\doibase 10.1209/0295-5075/110/67003} {\bibfield  {journal} {\bibinfo
  {journal} {Europhysics Letters}\ }\textbf {\bibinfo {volume} {110}},\
  \bibinfo {pages} {67003} (\bibinfo {year} {2015})}\BibitemShut {NoStop}%
\bibitem [{\citenamefont {Bruch}\ \emph {et~al.}(2016)\citenamefont {Bruch},
  \citenamefont {Thomas}, \citenamefont {Viola~Kusminskiy}, \citenamefont {von
  Oppen},\ and\ \citenamefont {Nitzan}}]{bruch2016a}%
  \BibitemOpen
  \bibfield  {author} {\bibinfo {author} {\bibfnamefont {A.}~\bibnamefont
  {Bruch}}, \bibinfo {author} {\bibfnamefont {M.}~\bibnamefont {Thomas}},
  \bibinfo {author} {\bibfnamefont {S.}~\bibnamefont {Viola~Kusminskiy}},
  \bibinfo {author} {\bibfnamefont {F.}~\bibnamefont {von Oppen}}, \ and\
  \bibinfo {author} {\bibfnamefont {A.}~\bibnamefont {Nitzan}},\ }\href
  {\doibase 10.1103/PhysRevB.93.115318} {\bibfield  {journal} {\bibinfo
  {journal} {Phys. Rev. B}\ }\textbf {\bibinfo {volume} {93}},\ \bibinfo
  {pages} {115318} (\bibinfo {year} {2016})}\BibitemShut {NoStop}%
\bibitem [{\citenamefont {Baines}\ \emph {et~al.}(2012)\citenamefont {Baines},
  \citenamefont {Meunier}, \citenamefont {Mailly}, \citenamefont {Wieck},
  \citenamefont {B\"auerle}, \citenamefont {Saminadayar}, \citenamefont
  {Cornaglia}, \citenamefont {Usaj}, \citenamefont {Balseiro},\ and\
  \citenamefont {Feinberg}}]{baines2012a}%
  \BibitemOpen
  \bibfield  {author} {\bibinfo {author} {\bibfnamefont {D.~Y.}\ \bibnamefont
  {Baines}}, \bibinfo {author} {\bibfnamefont {T.}~\bibnamefont {Meunier}},
  \bibinfo {author} {\bibfnamefont {D.}~\bibnamefont {Mailly}}, \bibinfo
  {author} {\bibfnamefont {A.~D.}\ \bibnamefont {Wieck}}, \bibinfo {author}
  {\bibfnamefont {C.}~\bibnamefont {B\"auerle}}, \bibinfo {author}
  {\bibfnamefont {L.}~\bibnamefont {Saminadayar}}, \bibinfo {author}
  {\bibfnamefont {P.~S.}\ \bibnamefont {Cornaglia}}, \bibinfo {author}
  {\bibfnamefont {G.}~\bibnamefont {Usaj}}, \bibinfo {author} {\bibfnamefont
  {C.~A.}\ \bibnamefont {Balseiro}}, \ and\ \bibinfo {author} {\bibfnamefont
  {D.}~\bibnamefont {Feinberg}},\ }\href {\doibase 10.1103/PhysRevB.85.195117}
  {\bibfield  {journal} {\bibinfo  {journal} {Phys. Rev. B}\ }\textbf {\bibinfo
  {volume} {85}},\ \bibinfo {pages} {195117} (\bibinfo {year}
  {2012})}\BibitemShut {NoStop}%
\bibitem [{\citenamefont {Hensgens}\ \emph {et~al.}(2017)\citenamefont
  {Hensgens}, \citenamefont {Fujita}, \citenamefont {Janssen}, \citenamefont
  {Li}, \citenamefont {Diepen}, \citenamefont {Reichl}, \citenamefont
  {Wegscheider}, \citenamefont {Sarma},\ and\ \citenamefont
  {Vandersypen}}]{hensgens2017a}%
  \BibitemOpen
  \bibfield  {author} {\bibinfo {author} {\bibfnamefont {T.}~\bibnamefont
  {Hensgens}}, \bibinfo {author} {\bibfnamefont {T.}~\bibnamefont {Fujita}},
  \bibinfo {author} {\bibfnamefont {L.}~\bibnamefont {Janssen}}, \bibinfo
  {author} {\bibfnamefont {X.}~\bibnamefont {Li}}, \bibinfo {author}
  {\bibfnamefont {C.~J.~V.}\ \bibnamefont {Diepen}}, \bibinfo {author}
  {\bibfnamefont {C.}~\bibnamefont {Reichl}}, \bibinfo {author} {\bibfnamefont
  {W.}~\bibnamefont {Wegscheider}}, \bibinfo {author} {\bibfnamefont {S.~D.}\
  \bibnamefont {Sarma}}, \ and\ \bibinfo {author} {\bibfnamefont {L.~M.~K.}\
  \bibnamefont {Vandersypen}},\ }\href {\doibase doi:10.1038/nature23022}
  {\bibfield  {journal} {\bibinfo  {journal} {Nature}\ }\textbf {\bibinfo
  {volume} {548}},\ \bibinfo {pages} {70} (\bibinfo {year} {2017})}\BibitemShut
  {NoStop}%
\bibitem [{\citenamefont {Bayer}\ \emph {et~al.}(2017)\citenamefont {Bayer},
  \citenamefont {Wagner}, \citenamefont {Rugeramigabo},\ and\ \citenamefont
  {Haug}}]{bayer2017a}%
  \BibitemOpen
  \bibfield  {author} {\bibinfo {author} {\bibfnamefont {J.~C.}\ \bibnamefont
  {Bayer}}, \bibinfo {author} {\bibfnamefont {T.}~\bibnamefont {Wagner}},
  \bibinfo {author} {\bibfnamefont {E.~P.}\ \bibnamefont {Rugeramigabo}}, \
  and\ \bibinfo {author} {\bibfnamefont {R.~J.}\ \bibnamefont {Haug}},\ }\href
  {\doibase 10.1103/PhysRevB.96.235305} {\bibfield  {journal} {\bibinfo
  {journal} {Phys. Rev. B}\ }\textbf {\bibinfo {volume} {96}},\ \bibinfo
  {pages} {235305} (\bibinfo {year} {2017})}\BibitemShut {NoStop}%
\bibitem [{\citenamefont {den Broeck}(2005)}]{vandenbroeck2005a}%
  \BibitemOpen
  \bibfield  {author} {\bibinfo {author} {\bibfnamefont {C.~V.}\ \bibnamefont
  {den Broeck}},\ }\href@noop {} {\bibfield  {journal} {\bibinfo  {journal}
  {Physical Review Letters}\ }\textbf {\bibinfo {volume} {95}},\ \bibinfo
  {pages} {190602} (\bibinfo {year} {2005})}\BibitemShut {NoStop}%
\bibitem [{\citenamefont {Gomez-Marin}\ and\ \citenamefont
  {Sancho}(2006)}]{gomez_marin2006a}%
  \BibitemOpen
  \bibfield  {author} {\bibinfo {author} {\bibfnamefont {A.}~\bibnamefont
  {Gomez-Marin}}\ and\ \bibinfo {author} {\bibfnamefont {J.~M.}\ \bibnamefont
  {Sancho}},\ }\href {\doibase 10.1103/PhysRevE.74.062102} {\bibfield
  {journal} {\bibinfo  {journal} {Phys. Rev. E}\ }\textbf {\bibinfo {volume}
  {74}},\ \bibinfo {pages} {062102} (\bibinfo {year} {2006})}\BibitemShut
  {NoStop}%
\bibitem [{\citenamefont {Sheng}\ and\ \citenamefont {Tu}(2013)}]{sheng2013a}%
  \BibitemOpen
  \bibfield  {author} {\bibinfo {author} {\bibfnamefont {S.}~\bibnamefont
  {Sheng}}\ and\ \bibinfo {author} {\bibfnamefont {Z.~C.}\ \bibnamefont {Tu}},\
  }\href {\doibase 10.1088/1751-8113/46/40/402001} {\bibfield  {journal}
  {\bibinfo  {journal} {Journal of Physics A: Mathematical and Theoretical}\
  }\textbf {\bibinfo {volume} {46}},\ \bibinfo {pages} {402001} (\bibinfo
  {year} {2013})}\BibitemShut {NoStop}%
\end{thebibliography}
%


\appendix



\section[Heisenberg equations]{Heisenberg equations for the phonon mapping}\label{APP:phonon_phonon}

The Heisenberg equations of motion for a system observable $A=A^\dagger$ read in the original representation
\begin{align}
\dot{A} &= \ii S_1(t) + \ii S_2(t) \sum_k \left(h_k a_k + h_k^* a_k^\dagger\right)\,,\qquad S_1(t) = [H_S, A]\,,\qquad S_2(t) = [S, A]\,,\nn
\dot{a}_k &= -\ii \omega_k a_k -\ii h_k^* S\,,\qquad 
\dot{a}_k^\dagger = +\ii \omega_k a_k^\dagger + \ii h_k S\,.
\end{align}
We now Fourier-transform these equations according to $\int [\ldots] e^{+\ii z t} dt$ with the convention $\Im z > 0$.
In $z$-space, the creation and annihilation operators are no longer adjoint to each other, but we will keep the $\dagger$-notation.
This yields the algebraic equations (convolution theorem)
\begin{align}
\ii z A(z) &= \ii S_1(z) + \frac{\ii}{2\pi} \int S_2(z') \sum_k \left[h_k a_k(z-z') + h_k^* a_k^\dagger(z-z')\right] dz'\,,\nn
\ii z a_k(z) &= -\ii \omega_k a_k(z) -\ii h_k^* S(z)\,,\qquad
\ii z a_k^\dagger(z) = +\ii \omega_k a_k^\dagger(z) + \ii h_k S(z)\,.
\end{align}
We can solve the last two equations $a_k(z) = -\frac{h_k^*}{z+\omega_k} S(z)$ and $a_k^\dagger(z) = +\frac{h_k}{z-\omega_k} S(z)$, and
insert them into the first
\begin{align}
z A(z) &= S_1(z)+ \frac{1}{2\pi} \int S_2(z') \left[\sum_k \frac{-\abs{h_k}^2}{z-z'+\omega_k} + \sum_k \frac{+\abs{h_k}^2}{z-z'-\omega_k}\right] S(z-z') dz'\nn
&= S_1(z) + \frac{1}{2\pi} \int S_2(z') \left[\frac{1}{\pi} \int_0^\infty \frac{\omega \Gamma^{(0)}(\omega)}{(z-z')^2-\omega^2} d\omega\right] S(z-z') dz'\nn
&= S_1(z) - \frac{1}{2\pi} \int S_2(z') \frac{1}{2} W^{(0)}(z-z') S(z-z') dz'\,.
\end{align}
Here, we have in the first step used the fact that the harmonic oscillator frequencies $\omega_k$ are by construction all positive and we have introduced the Cauchy transform
\begin{align}
W^{(n)}(z) = \frac{2}{\pi} \int_0^\infty \frac{\omega \Gamma^{(n)}(\omega)}{\omega^2-z^2} d\omega = \frac{1}{\pi} \int_{-\infty}^{+\infty} \frac{\Gamma^{(n)}(\omega)}{\omega-z} d\omega\,,
\end{align}
where the last equality sign holds for analytic continuation as an odd function $\Gamma(-\omega)=-\Gamma(+\omega)$.
In particular, we note the important property
\begin{align}
\Gamma^{(n)}(\omega) = \lim_{\epsilon\to 0^+} \Im W^{(n)}(\omega+\ii\epsilon)\,.
\end{align}

Similarly, we can derive the Heisenberg equations of motion in the mapped representation, and Fourier-transform them according to the
same prescription, yielding
\begin{align}
z A(z) &= S_1(z) + \frac{\lambda}{2\pi} \int S_2(z') \left[b(z-z') + b^\dagger(z-z')\right] dz'\,,\nn
z b(z) &= -\lambda S(z) - \Omega b(z) - \sum_k \left[H_k b_k(z) + H_k^* b_k^\dagger(z)\right]\,,\nn
z b^\dagger(z) &= +\lambda S(z) + \Omega b^\dagger(z) + \sum_k \left[H_k b_k(z) + H_k^* b_k^\dagger(z)\right]\,,\nn
z b_k(z) &= -\Omega_k b_k(z) - H_k^* \left[b(z) + b^\dagger(z)\right]\,,\qquad
z b_k^\dagger(z) = +\Omega_k b_k^\dagger(z) + H_k \left[b(z) + b^\dagger(z)\right]\,.
\end{align}
Again, we follow the approach of successively eliminating the $b_k(z)$, $b_k^\dagger(z)$, and then the 
$b(z)$, $b^\dagger(z)$ variables, yielding for the remaining equation
\begin{align}
z A(z) = S_1(z) + \frac{1}{2\pi} \int S_2(z') \frac{2 \lambda^2 \Omega}{(z-z')^2 - \Omega^2 + \Omega W^{(1)}(z-z')} S(z-z') dz'\,.
\end{align}
Comparing this with the original representation, we can infer a relation between $W^{(0)}(z)$ and $W^{(1)}(z)$, which
can be used to obtain the transformed spectral density
\begin{align}
\Gamma^{(1)}(\omega) &= -\lim_{\epsilon\to 0^+} \Im \frac{4\lambda^2}{W^{(0)}(\omega+\ii\epsilon)}
= \frac{+4 \lambda^2 \Gamma^{(0)}(\omega)}
{\left[\frac{1}{\pi}{\cal P} \int \frac{\Gamma^{(0)}(\omega')}{\omega-\omega'} d\omega'\right]^2
+ \left[\Gamma^{(0)}(\omega)\right]^2}\,.
\end{align}


\section{Heisenberg equations for the particle mapping}\label{APP:particle_particle}

Now, the Heisenberg equations of motion for a system observable $A=A^\dagger$ read in the original representation
\begin{align}
\dot{A} &= \ii S_1(t) + \ii S_2(t) \sum_k h_k^* a_k^\dagger - \ii S_2^\dagger(t) \sum_k h_k a_k\,,\qquad S_1(t) = [H_S, A]\,,\qquad S_2(t) = [S, A]\,,\nn
\dot{a}_k &= -\ii \omega_k a_k -\ii h_k^* S\,,\qquad 
\dot{a}_k^\dagger = +\ii \omega_k a_k^\dagger + \ii h_k S^\dagger\,.
\end{align}
Fourier-transformation yields 
\begin{align}
z A(z) &= S_1(z) + \frac{1}{2\pi} \int \left[S_2(z') \sum_k h_k^* a_k^\dagger(z-z') - S_2^\dagger(z') \sum_k h_k a_k(z-z')\right] dz'\,,\nn
z a_k(z) &= -\omega_k a_k(z) - h_k^* S(z)\,,\qquad
z a_k^\dagger(z) = +\omega_k a_k^\dagger(z) + h_k S^\dagger(z)\,.
\end{align}
Inserting the solutions of the last two equations into the first we get
\begin{align}
z A(z) = S_1(z) + \frac{1}{2\pi} \int \left[S_2(z') \sum_k \frac{\abs{h_k}^2}{z-z'-\omega_k} S^\dagger(z-z') + S_2^\dagger(z') \sum_k \frac{\abs{h_k}^2}{z-z'+\omega_k} S(z-z')\right]\,.
\end{align}

In the mapped representation, we have
\begin{align}
\dot{A} &= \ii S_1(t) + \ii \lambda S_2(t) b^\dagger - \ii \lambda S_2^\dagger(t) b\,,\nn
\dot{b} &= -\ii \lambda S - \ii \Omega b - \ii \sum_k H_k b_k\,,\qquad
\dot{b}^\dagger = +\ii\lambda S^\dagger + \ii \Omega b^\dagger + \ii \sum_k H_k^* b_k^\dagger\,,\nn
\dot{b}_k &= -\ii H_k^* b-\ii\Omega_k b_k\,,\qquad
\dot{b}_k^\dagger = +\ii H_k b^\dagger + \ii \Omega_k b_k^\dagger\,,
\end{align}
such that Fourier transformation yields
\begin{align}
z A(z) &= S_1(z) + \frac{\lambda}{2\pi} \int\left[S_2(z') b^\dagger(z-z') - S_2^\dagger(z') b(z-z')\right] dz'\,,\nn
z b(z) &= - \lambda S(z) - \Omega b(z) - \sum_k H_k b_k(z)\,,\qquad
z b^\dagger(z) = +\lambda S^\dagger(z) + \Omega b^\dagger(z) + \sum_k H_k^* b_k^\dagger(z)\,,\nn
z b_k(z) &= -H_k^* b(z) - \Omega_k b_k(z)\,,\qquad
z b_k^\dagger(z) = +H_k b^\dagger(z) + \Omega_k b_k^\dagger(z)\,.
\end{align}
Successive elimination of the last four equations yields for the remaining one
\begin{align}
z A(z) &= S_1(z)\\
&+ \frac{\lambda}{2\pi} \int\left[ S_2(z') \frac{+\lambda}{z-z'-\Omega-\sum_k \frac{\abs{H_k}^2}{z-z' -\Omega_k}} S^\dagger(z-z')
+ S_2^\dagger(z') \frac{\lambda}{z-z'+\Omega-\sum_k \frac{\abs{H_k}^2}{z-z' +\Omega_k}} S(z-z')\right]\,.\nonumber
\end{align}
From comparison with the first representation, we conclude
\begin{align}
\sum_k \frac{\abs{h_k}^2}{z-\omega_k} = \frac{\lambda^2}{z-\Omega-\sum_k \frac{\abs{H_k}^2}{z-\Omega_k}}\,,\qquad
\sum_k \frac{\abs{h_k}^2}{z+\omega_k} = \frac{\lambda^2}{z+\Omega-\sum_k \frac{\abs{H_k}^2}{z+\Omega_k}}\,,
\end{align}
where the second equation just encodes the first at $-z$ and is therefore not independent.

From realizing that 
\begin{align}
\lim_{\epsilon\to 0} \frac{1}{2\pi} \int_0^\infty \frac{\Gamma(\omega')}{\omega-\omega'+\ii\epsilon} d\omega' \stackrel{\omega>0}{=} 
\frac{1}{2\pi} {\cal P} \int_0^\infty \frac{\Gamma(\omega')}{\omega-\omega'} d\omega' - \frac{\ii}{2} \Gamma(\omega)\,,
\end{align}
we can use e.g. the first of these relations to infer a mapping relation between the spectral densities,
\begin{align}
\Gamma^{(1)}(\omega) = \frac{4 \lambda^2 \Gamma^{(0)}(\omega)}
{\left[\frac{1}{\pi}{\cal P}\int_0^\infty \frac{\Gamma^{(0)}(\omega')}{\omega-\omega'} d\omega'\right]^2 + \left[\Gamma^{(0)}(\omega)\right]^2}\,,
\end{align}
where $\omega>0$ is assumed throughout.


\section{Heisenberg equations for fermionic reservoirs}\label{APP:fermion_fermion}

To avoid case distinctions on whether the system operator $A$ commutes or anti-commutes with the coupling operator, we just consider the Heisenberg equations for the creation
and annihilation operators.
In the original representation, they become
\begin{align}
\dot{c} = \ii [H_S, c] + \ii \sum_k t_k c_k = \ii S(t) + \ii \sum_k t_k c_k\,,\qquad
\dot{c}_k = \ii t_k^* c - \ii \epsilon_k c_k\,,
\end{align}
and similarly for the creation operators. 
Since at this level they do not mix, we consider only the annihilation operators.
Fourier-transformation yields
\begin{align}
z c(z) = S(z) + \sum_k t_k c_k(z)\,,\qquad
z c_k(z) = t_k^* c(z) - \epsilon_k c_k(z)\,.
\end{align}
Eliminating the second equation then gives
\begin{align}
z c(z) = S_1(z) + \sum_k \frac{\abs{t_k}^2}{z+\epsilon_k} c(z)\,.
\end{align}

In contrast, the mapped representation yields
\begin{align}
\dot{c} = \ii S(t) + \ii \lambda d\,,\qquad
\dot{d} = -\ii \lambda c - \ii \epsilon d + \ii \sum_k T_k d_k\,,\qquad
\dot{d}_k = \ii T_k^* d - \ii \epsilon_k d_k\,.
\end{align}
Fourier-transforming and eliminating the non-system variables then gives
\begin{align}
z c(z) = S(z) - \frac{\lambda^2}{z+\epsilon-\sum_k \frac{\abs{T_k}^2}{z+\epsilon_k}} c(z)\,,
\end{align}
and from comparison we get the relation
\begin{align}
\sum_k \frac{\abs{t_k}^2}{z+\epsilon_k} = - \frac{\lambda^2}{z+\epsilon-\sum_k \frac{\abs{T_k}^2}{z+\epsilon_k}}\,.
\end{align}
Converting the sums to integrals and evaluating at $z=-\omega+\ii\delta$ when $\delta\to 0^+$ we obtain a mapping relation between the fermionic spectral densities:
\begin{align}
\Gamma^{(1)}(\omega) = \frac{4 \lambda^2 \Gamma^{(0)}(\omega)}{\left[\frac{1}{\pi} {\cal P}\int \frac{\Gamma^{(0)}(\omega')}{\omega-\omega'} d\omega'\right]^2 + \left[\Gamma^{(0)}(\omega)\right]^2}\,.
\end{align}


\end{document}